\documentstyle[preprint,prl,aps,epsfig]{revtex}
\tightenlines
\begin{document}

\title{Jet Photoproduction in Peripheral Heavy-Ion
Collisions}

\author{R. Vogt} 
\address{Nuclear Science Division,
Lawrence Berkeley National Laboratory, Berkeley, CA, 94720\break and \break
Physics Department, University of California, Davis, CA, 95616}
\maketitle

\begin{abstract}
In peripheral relativistic heavy ion collisions at the Large Hadron Collider, 
jet$+$jet and photon$+$jet final states can be produced 
when a photon from the
virtual photon field surrounding the nucleus interacts with a parton in the
opposite nucleus (direct production).  The virtual photon may also fluctuate
into states with multiple gluons and $q \overline q$ pairs (resolved 
production), opening more
channels for jet photoproduction.  We compare the rates for direct and resolved
jet$+$jet and photon$+$jet production to explore the sensitivity to the nuclear
and photon parton distribution functions.  We calculate the transverse momentum
distributions of both partonic jets and leading hadrons 
produced by jet fragmentation.
\end{abstract}

\section{Introduction}

We discuss  measurements of the nuclear parton distributions
via dijet (jet$+$jet) and $\gamma+$jet 
photoproduction in ultraperipheral heavy ion collisions at the Large Hadron
Collider (LHC).  
In these collisions, the accelerated ion is surrounded
by a cloud of almost real photons of virtuality $|q^2| < (\hbar c/R_A)^2$
where $R_A$ is the nuclear radius.  The virtuality, less than $(60 \, {\rm
MeV})^2$ for nuclei with $A> 16$, can be neglected.  Since photon interactions
are long range, they can interact with partons in the opposite
nucleus even when the nuclei themselves do not interpenetrate.  Because the
photon energies are less than those of the nucleons, these photonuclear
interactions have a smaller average center of mass energy than hadronic
parton-parton collisions.  However, even though the energy is smaller, the
coherent photon beams have a flux proportional to the square of the nuclear 
charge, $Z$, enhancing the rates relative to those of photoproduction in
$pp$ collisions at the same energy.  Thus photoproduction rates in heavy ion
collisions can be high.

Photoproduction occurs two ways in heavy ion collisions: ``direct'' and
``resolved'' production.   
``Direct'' photoproduction occurs when a photon emitted from one
nucleus interacts with a parton from the other nucleus.
The photon can also fluctuate into states with multiple $q\overline
q$ pairs and gluons, {\it i.e.}\ $|n(q \overline q)m(g)\rangle$. One
of these photon components can interact with a quark or gluon from the
target nucleus (``resolved'' production) \cite{Witten:ju}.  
The photon components are
described by parton densities similar to those used for protons except
that no useful momentum sum rule applies to the photon 
\cite{Sjostrand:1996wz}.
The quark and gluon constituents of the photon open up more channels for 
jet photoproduction and could, in principle, lead to larger
rates for resolved production in certain regions of phase space.

In central collisions at RHIC, leading particles in jets are 
easier to detect above the high charged particle multiplicity background than
the jets themselves since these high transverse momentum, $p_T$,
particles can be tracked through the detector \cite{Adler:2002tq,Chiu:2002ma}.
In peripheral collisions, especially at LHC energies, jets should be easier
to isolate and may be observed directly using standard high energy jet 
detection
techniques.  We thus discuss the $p_T$ distributions of both jets
and leading particles.  We work at leading order, LO, 
to avoid any ambiguities such as
jet reconstruction and cone size.  The ratio of 
the next-to-leading order, NLO, to LO jet cross sections appears to be 
relatively constant at high $p_T$ \cite{Eskola:cj}.
We discuss the fragmentation of
jets and present the transverse momentum distributions of 
charged pions, charged kaons and protons/antiprotons.

\section{Dijet production}

\subsection{Direct dijet production}

\begin{figure}[htbp]
\setlength{\epsfxsize=0.95\textwidth}
\setlength{\epsfysize=0.5\textheight}
\centerline{\epsffile{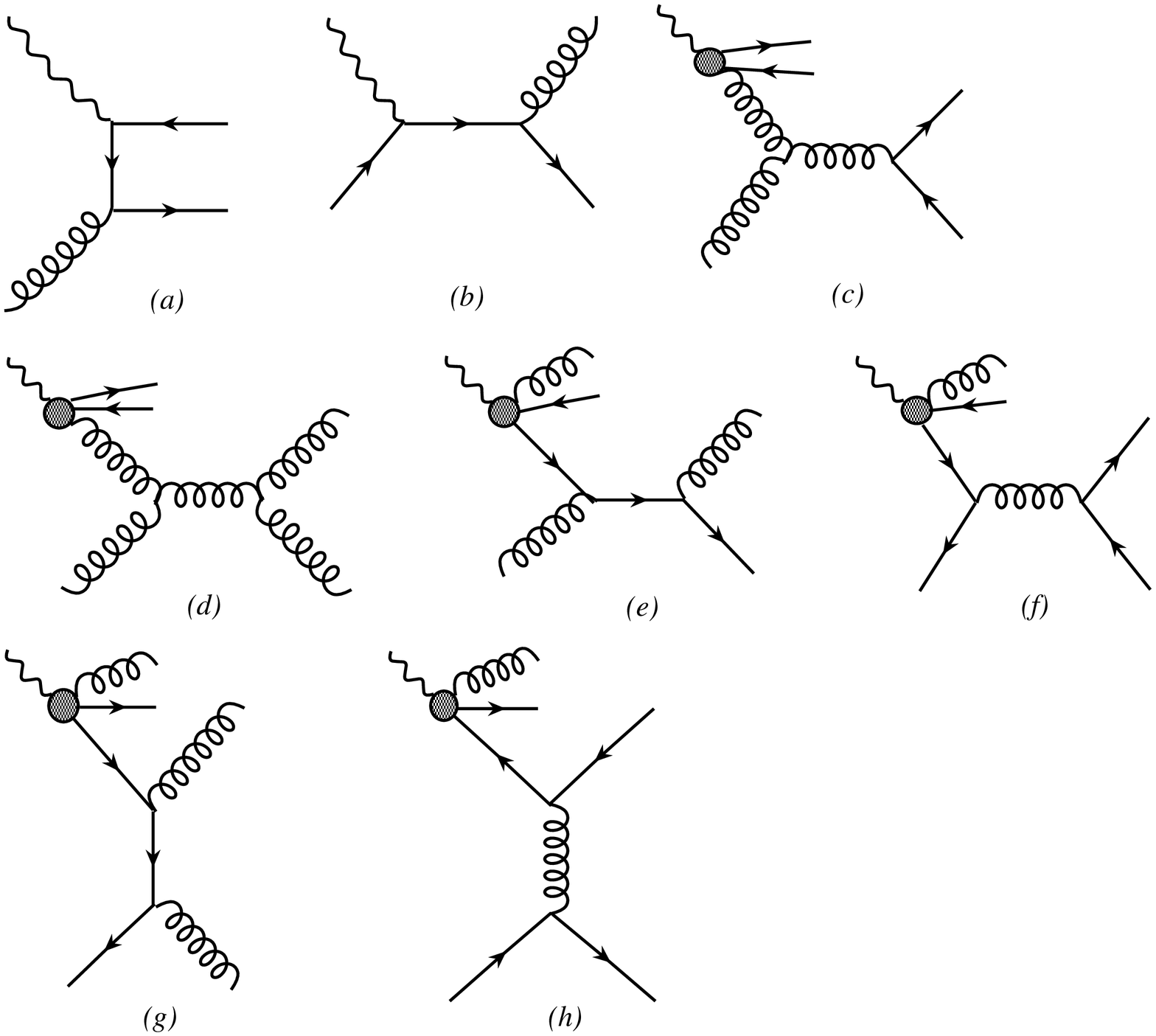}}
\caption[]{Feynman diagrams for dijet photoproduction from direct, (a) and (b),
and resolved photons, (c)-(h).  Only a sample of the resolved diagrams are
shown. Crossed diagrams are not shown.}
\label{dijetdia}
\end{figure}

The first hadronic reaction we study is $\gamma(k) + N(P_2) \rightarrow {\rm
jet}(p_1) \, 
+ \, {\rm jet}(p_2) \, + \, X$ where $k$ and $P_2$ are the photon and 
nucleon four-momenta.  The diagrams for direct and resolved dijet
hadroproduction are given in Fig.~\ref{dijetdia}.
The two parton-level
contributions to the jet yield in direct photoproduction 
are $\gamma(k) + g(x_2P_2) \rightarrow q(p_1) + \overline q(p_2)$ 
and $\gamma(k) + q(x_2P_2) \rightarrow g(p_1) + q(p_2)$ where also $q
\rightarrow \overline q$.  The produced partons are 
massless, requiring a minimum $p_T$ to keep the cross section finite.  
At LO, the jet yield is equivalent to the high $p_T$ parton yield.
The jet $p_T$ distribution is modified for photoproduction from {\it e.g.} 
Refs.~\cite{Emel'yanov:1999bn,Eskola:1988yh,Eskola:1996ce},  
\begin{eqnarray}
\label{jetdireq} S_{NN}^2
\frac{d^2\sigma_{\gamma A \rightarrow {\rm jet} \, + \, {\rm jet}+ X}^{\rm 
dir}}{dT dU d^2b} = 
2 \int dz \int_{k_{\rm min}}^\infty dk \frac{d^3N_\gamma}{dk d^2b}
\int_{x_{2_{\rm min}}}^1 \frac{dx_2}{x_2} \bigg[ \sum_{i,j,l =
q, \overline q, g} F_i^A(x_2,\mu^2, \vec b, z) s^2 \frac{d^2\sigma_{\gamma i 
\rightarrow jl}}{dt du} \bigg]
\, \,  ,
\end{eqnarray}
where $x_2$ is the fraction of the initial hadron momentum carried by the 
interacting parton and $\mu$ is the momentum scale of the interaction.
Recall that the extra factor of two on the right-hand side of
Eq.~(\ref{jetdireq}) arises because 
both nuclei can serve as photon sources in $AA$ collisions.  The partonic
cross sections are
\begin{eqnarray}
s^2 \frac{d^2 \sigma_{\gamma g \rightarrow q \overline q}}{dt du} & = 
& \pi \alpha_s(\mu^2) \alpha e_Q^2  
\bigg[\frac{t^2 +u^2}{t u}  \bigg] \delta(s + t + u) \, \, , \label{gamApart}
\\ 
s^2 \frac{d^2\sigma_{\gamma q \rightarrow gq}}{dt du} & = & - \frac{8}{3} \pi 
\alpha_s(\mu^2)
\alpha e_Q^2 \bigg[ \frac{s^2 + t^2}{s t}\bigg] \delta(s + t + u) \, \, .
\label{QCDcompt}
\end{eqnarray}
The first is the photon-gluon fusion cross section, the only contribution to
massive
$Q \overline Q$ photoproduction \cite{Klein:2002wm}, while the second is the
QCD Compton process.
At LO, the partonic cross section is 
proportional to $\alpha \alpha_s(\mu^2) e_Q^2$, where
$\alpha_s(\mu^2)$ is the strong coupling constant to one loop, 
$\alpha=e^2/\hbar c$ is the
electromagnetic coupling constant, and $e_Q$ is the light quark charge, $e_u = 
e_c = 2/3$
and $e_d = e_s = -1/3$. The partonic invariants, $s$,
$t$, and $u$, are defined as $s = (k + x_2 P_2)^2$, $t = (k -
p_1)^2 = (x_2 P_2 - p_2)^2$, and $u = (x_2 P_2 -
p_1)^2 = (k - p_2)^2$.  In this case, $s = 4k \gamma_L
x_2m_p$ where $\gamma_L$ is the Lorentz boost of a single beam and
$m_p$ is the proton mass.  Since $k$ can be a continuum of
energies up to $E_{\rm beam} = \gamma_L m_p$, we define $x_1 = k/P_1$,
analogous to the parton momentum fraction in the nucleon
where $P_1$ is the nucleon
four momentum. For a detected parton in a nucleon-nucleon collision, the
hadronic invariants are then $S_{NN} = (P_1 + P_2)^2$, $T = (P_2 - p_1)^2$,
and $U = (P_1 - p_1)^2$.

The produced parton rapidities are $y_1$ and $y_2$.
The jet parton rapidity is related to the invariant $T$ by $T = -
\sqrt{S_{NN}} p_T e^{-y_1}$.  At LO, $x_1
= (p_T/\sqrt{S_{NN}})(e^{y_1} + e^{y_2})$ and $x_2 =
(p_T/\sqrt{S_{NN}})(e^{-y_1} + e^{-y_2})$.  We calculate $x_1$ and $x_2$ as
in an $NN$ collision and then determine the flux in the lab frame for
$k = x_1 \gamma_L m_p$, equivalent to the center of mass frame in a
collider.  The photon flux is exponentially suppressed for $k>\gamma_L
\hbar c/R_A$, corresponding to a momentum fraction $x_1 > \hbar
c/m_pR_A$.  The maximum $\gamma N$ center of mass energy,
$\sqrt{S_{\gamma N}}$, is much lower than the hadronic $\sqrt{S_{NN}}$.  
The equivalent hadronic invariants can be defined for photon four
momentum $k$ as $S_{\gamma N} = (k + P_2)^2$, $T_{\gamma N} = (P_2 -
p_1)^2$, and $U_{\gamma N} = (k - p_1)^2$
\cite{Smith:1991pw}.  The partonic and equivalent hadronic invariants for fixed
$k$ are related by $s = x_2S_{\gamma N}$, $t = U_{\gamma N}$, and
$u = x_2 T_{\gamma N}$.

The photon flux is given by the Weizs\"acker-Williams method.  The
flux from a charge $Z$ nucleus at a distance $r$ is
\begin{equation}
{d^3N_\gamma \over dkd^2r} = 
{Z^2\alpha w^2\over \pi^2kr^2} \left[ K_1^2(w) + {1\over
\gamma_L^2} K_0^2(w) \right] \, \, 
\label{wwr}
\end{equation}
where $w=kr/\gamma_L$ and $K_0(w)$ and $K_1(w)$ are modified Bessel
functions.  The photon flux decreases exponentially above a cutoff
energy determined by the nuclear size.  In the lab frame, the
cutoff is $k_{\rm max} \approx \gamma_L \hbar c/R_A$.  In the rest frame of the
target nucleus, the cutoff is boosted to 
$E_{\rm max}=(2\gamma_L^2-1)\hbar c/R_A$.  Table~\ref{gamfacs} shows, for $AA$
collisions at the LHC, the
nucleon-nucleon center of mass energies, $\sqrt{S_{NN}}$, the
beam energies, $E_{\rm beam}$, Lorentz factors, $\gamma_L$, maximum photon
energies in the lab and rest frames, $k_{\rm
max}$ and $E_{\rm max}$ respectively, 
as well as the corresponding maximum $\gamma p \rightarrow q \overline q$ 
center of mass energy,
$\sqrt{S_{\gamma N}} = \sqrt{2E_{\rm max}m_p}$, for single
photon interactions with protons.   
We have also included the $AA$ luminosities in
Table~\ref{gamfacs} to aid the calculation of rates.  

The total photon flux striking the target nucleus is the integral of
Eq.~(\ref{wwr}) over the transverse area of the target for all impact
parameters subject to the constraint that the two nuclei do not
interact hadronically \cite{Klein:1999qj}.  This must be calculated
numerically.  However, an analytic approximation for $AA$
collisions is given by the photon flux integrated over distances $r>2R_A$,
\begin{equation}
{dN_\gamma\over dk} = 
{2Z^2 \alpha \over\pi k} \left[ w_R K_0(w_R)
K_1(w_R)- {w_R^2\over 2} \big(K_1^2(w_R)-K_0^2(w_R)
\big) \right] \, \, 
\label{analflux}
\end{equation}
where $w_R = 2kR_A/\gamma_L$.
The difference between the numerical calculation and the analytic
results is typically less than 15\% except for photon energies
near the cutoff. We use the more accurate numerical
calculations.  

The nuclear parton densities $F_i^A(x,\mu^2,\vec{b},z)$ can be
factorized into $x$ and $\mu^2$ independent nuclear density
distributions, position and nuclear-number independent nucleon parton
densities, and a shadowing function $S^i_A(x,\mu^2,\vec{b},z)$ that
describes the modification of the nuclear parton distributions in
position and momentum space.  Then 
\cite{Emel'yanov:1999bn,Emel'yanov:1997pu,Emel'yanov:1998df,Emel'yanov:1998sy,Vogt:2000hp}
\begin{eqnarray}
F_i^A(x,\mu^2,\vec{b},z) & = & \rho_A(\vec{b},z) S^i_A(x,\mu^2,\vec{b},z)
f_i^N(x,\mu^2) \label{nucglu} \, \, 
\end{eqnarray}
where $f^N_i(x,\mu^2)$ is the parton density in the nucleon.  We use 
the MRST LO parton distributions \cite{Martin:1998np} at $\mu^2= p_T^2$.  
In the absence of nuclear
modifications, $S^i \equiv1$.  The nuclear density
distribution, $\rho_A(\vec{b},z)$, is a Woods-Saxon
with parameters determined from electron
scattering data \cite{DeJager:1974dg}.  
Although most models of shadowing predict
a dependence on the parton position in the nucleus, 
we neglect any impact parameter
dependence here so that $S^i_A(x,\mu^2,\vec b,z) \rightarrow S^i_A(x,\mu^2)$.

The nuclear modification, collectively referred to as shadowing here, 
encompasses three $x$ regions: low $x$ shadowing where $S^i_A<1$, $x < 0.01$; 
antishadowing where $S^i_A>1$, $0.01 \leq x \leq 0.1$ and the EMC region
where again $S^i_A<1$ but $x \geq 0.2$. 
We use two parameterizations of nuclear shadowing 
which cover extremes of low $x$ gluon shadowing.  The Eskola
{\it et al.} parameterization, EKS98, is based on the GRV LO
\cite{Gluck:1991ng} parton densities.  At the minimum scale, $\mu_0$, valence 
quark shadowing is identical for
$u$ and $d$ quarks.  Likewise, the shadowing of $\overline u$ and
$\overline d$ quarks are identical at $\mu_0$. Although the light quark
shadowing ratios are not constrained to be equal at higher scales, the 
differences between them are small.  Shadowing of the heavier flavor
sea, $\overline s$ and higher, is calculated separately at $\mu_0$.  The
shadowing ratios for each parton type are evolved to LO for $1.5 < \mu <
100$ GeV and are valid for $x \geq 10^{-6}$ \cite{Eskola:1998iy,Eskola:1998df}.
Interpolation in nuclear mass number allows results to be obtained for
any input $A$.  The parameterization by Frankfurt, Guzey and
Strikman, FGS, combines Gribov theory with hard
diffraction \cite{Frankfurt:2003zd}.  It is based on the CTEQ5M 
\cite{Lai:1999wy} parton
densities and evolves each parton species separately to NLO for $2 < \mu
< 100$ GeV.  Although the given $x$ range is $10^{-5} < x < 0.95$, the
sea quark and gluon ratios are unity for $x > 0.2$ and $\mu_0$.  
The EKS98 valence quark shadowing ratios are used as input at $\mu_0$
since Gribov theory does not
predict valence shadowing.  The parameterization is available for four
different values of $A$: 16, 40, 110 and 206.  

\begin{figure}[htbp] 
\setlength{\epsfxsize=0.95\textwidth}
\setlength{\epsfysize=0.5\textheight}
\centerline{\epsffile{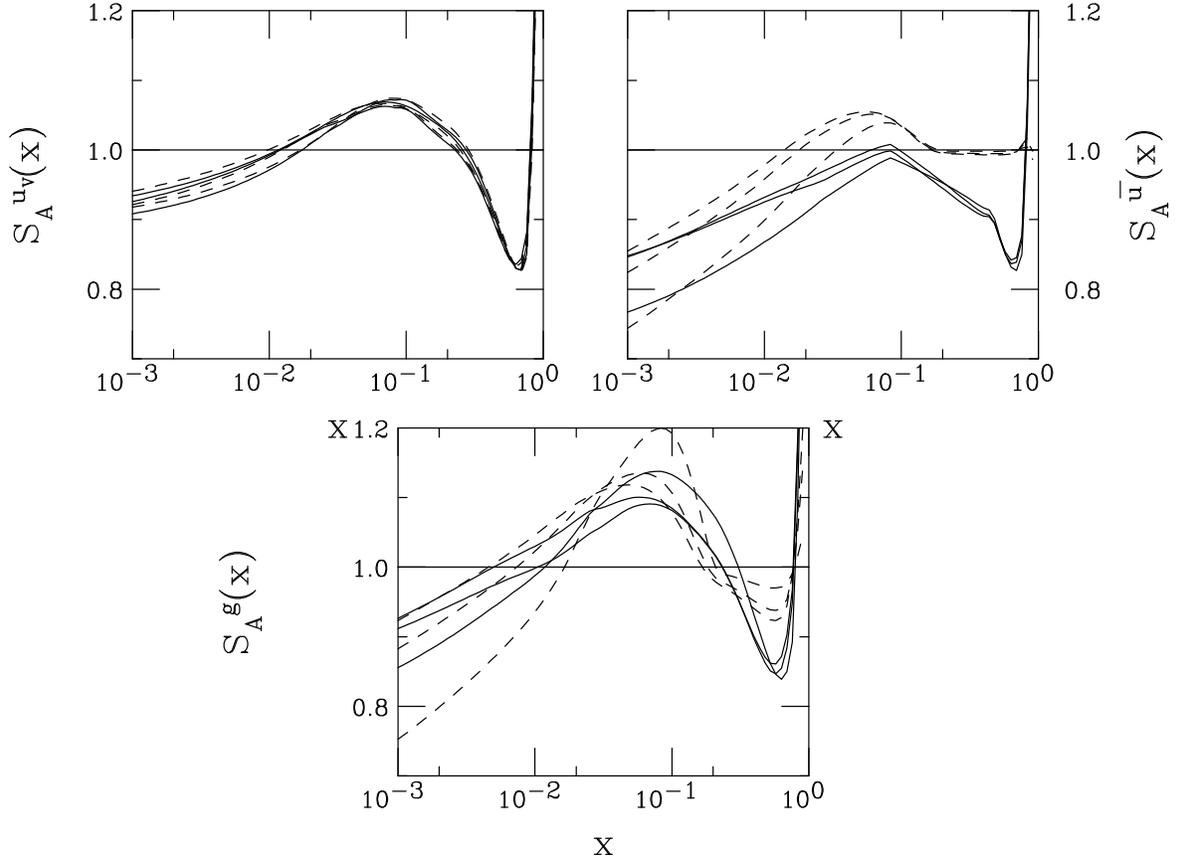}}
\caption[]{ The EKS98 and FGS shadowing parameterizations are compared at
the scale $\mu = 10$, 100 and 400 GeV for (a) valence quarks, (b) sea quarks
and (c) gluons.  The solid curves are the EKS98
parameterization, the dashed, FGS.  The lower values of $\mu$ give the lowest
values of $S_A^i$.
}
\label{shadcomp}
\end{figure}

Figure~\ref{shadcomp} compares
the two parameterizations for $A \approx 200$ and $\mu = 10$, 100 and 400 GeV.
Some care must be taken when applying these parameterizations to high $p_T$ 
since the upper limit of their fit range is 100 GeV. The 
valence ratios, identical for EKS98 and FGS at $\mu_0$, are somewhat different
at high scales because the evolution is different, LO for EKS98 and NLO for
FGS.  The sea quark ratios are rather dissimilar.  The EKS98
ratios have no antishadowing but, instead, show a peak at $S_A^{\overline u}
\approx 1$ when $x \approx 1$, decreasing smoothly at larger $x$, the EMC
region.  The FGS parameterization has an antishadowing peak at $x \approx
0.07$ after which $S_A^{\overline u} \approx 1$ for $x > 0.2$, as at $\mu_0$.
Thus quark shadowing in the EMC $x$ region will be stronger for EKS98.  At the
lowest $x$ and $\mu$ shown, the FGS gluon ratio shows stronger shadowing and
larger antishadowing than EKS98.  By $\mu \geq 100$ GeV, the results are
rather similar.  It is notable that at $\mu_0$ the FGS gluon ratio is 
$\approx 1$ for $x > 0.2$ and decreases below 1 at higher scales, unlike the
sea quark ratios, exhibiting an EMC-like effect.  At high $p_T$, therefore, the
EKS98 and FGS gluon ratios should not be significantly different.  As we have
already pointed out, although the upper limit on the range of both shadowing
parameterizations is 100 GeV, the parameterizations do not strongly
deviate from the trends at lower scales.  Thus the results can be taken
as indicative of the expected behavior.

There are
a few photon parton distributions
available~\cite{Gluck:1991jc,Gluck:1991ee,Drees:1984cx,Abramowicz:1991yb,Hagiwara:1994ag,Schuler:1995fk,Schuler:1996fc}.
The data \cite{PDFLIB,Bartel:1984cg} cannot definitively rule out any of 
these parton densities.  
As expected, $F_q^\gamma(x,\mu^2) = F_{\overline
q}^\gamma (x,\mu^2)$ flavor by flavor because there are 
no ``valence'' quarks in
the photon.  The gluon distribution in the photon is less well known.
We use the GRV-G LO set \cite{Gluck:1991jc,Gluck:1991ee}.  At $p_T>10$ GeV,
the difference in results due to the photon parton densities is minimal.

\begin{figure}[htbp]
\setlength{\epsfxsize=0.95\textwidth}
\setlength{\epsfysize=0.45\textheight}
\centerline{\epsffile{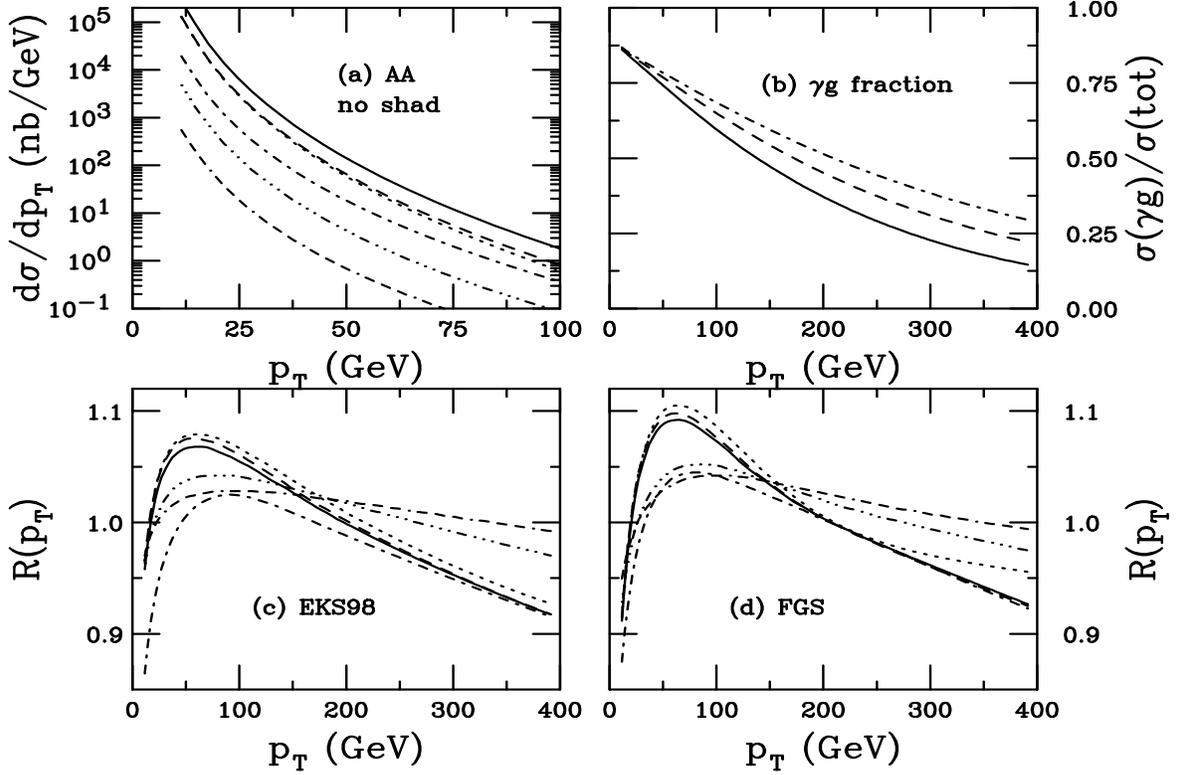}}
\caption[] {Direct jet photoproduction in peripheral collisions. (a)
The $p_T$ distributions for $|y_1| \leq 1$ are shown for $AA$ 
collisions.  The solid curves is the total for Pb ions while the produced 
quarks (dashed), antiquarks (dotted) and gluons (dot-dashed) are shown
separately.  The total production for Ar (dot-dot-dot-dashed) and O
(dot-dash-dash-dashed) ions are also shown. (b) The fraction of gluon-initiated
jets as a function of $p_T$ for Pb+Pb (solid), Ar+Ar (dashed) and O+O
(dot-dashed) interactions. (c) The EKS98 shadowing ratios for produced partons.
The solid curve is the total for Pb ions while the ratios for produced 
quarks (dashed), antiquarks (dotted) and gluons (dot-dashed) are shown
separately.  The total ratios for Ar (dot-dot-dot-dashed) and O
(dot-dash-dash-dashed) ions are also shown. (d) The same as (c) for FGS.}
\label{jetdir}
\end{figure}

The direct jet photoproduction $p_T$ distributions 
are given in Fig.~\ref{jetdir} for $AA$
interactions at the LHC.  For Pb+Pb collisions at $\sqrt{S_{NN}} = 5.5$ TeV, we
show the $p_T$ distributions of the produced quarks, 
antiquarks and gluons separately, as well as their sum.  
For Ar+Ar collisions at $\sqrt{S_{NN}}=6.3$ TeV and O+O collisions 
at $\sqrt{S_{NN}}=7$ TeV, we show only the total $p_T$ distributions.  
All the results are shown in the rapidity interval $|y_1|\leq 1$.

Quarks and antiquarks are produced in greatest abundance, with only a small
difference at high $p_T$.  Photon-gluon fusion alone
produces equal numbers of quarks and antiquarks.  The quark excess arises from
the QCD Compton diagram which also produces the small final-state 
gluon contribution.  The $\gamma (q + \overline q)$ 
contribution grows with $p_T$ since the valence quark distributions
eventually dominate production, 
as shown in Fig.~\ref{jetdir}(b) where the $\gamma g$
contribution is compared to the total.  At low $p_T$, the
$\gamma g$ contribution is $\approx 90$\% of the total, dropping to $10-30$\%
at $p_T \approx 400$ GeV.  At the large values of $x$ needed for high $p_T$ 
jet production, $f_p^{u_V} > f_p^g$.  
Thus the QCD Compton process eventually dominates dijet production,
albeit in a region of very low statistics.  The $\gamma g$ contribution is
larger for the lighter nuclei since the higher energies reduce the average $x$
values.  

The direct dijet photoproduction cross section
is significantly lower than the
hadroproduction cross section.  Some of this reduction is due to the different
couplings.  The photoproduction rate is reduced by a factor of $\alpha
e_Q^2/\alpha_s \approx 100$.  There are also fewer 
diagrams available for jet photoproduction relative to all $2
\rightarrow 2$ scatterings in hadroproduction.   In addition, the 
$gg \rightarrow gg$ hadroproduction process, with its high parton luminosity,
has no direct photoproduction equivalent.

Since the typical scales of jet production are large, 
the effects of shadowing, reflected in $R(p_T) = 
(d\sigma[S_A^i]/dp_T)/(d\sigma[S_A^i=1]/dp_T)$, 
are rather small, see Fig.~\ref{jetdir}(c) 
and (d), because the average $x$ is high.  The differences
between the two shadowing parameterizations are on the few percent level.
At low $p_T$, the produced quarks and antiquarks are mainly from gluons.
The produced gluons only come from quarks.  The peak for the produced quarks 
and antiquarks in $R(p_T)$
between $50 \leq p_T \leq 100$ GeV is
due to gluon antishadowing.
The total $R(p_T)$ for all produced partons in Pb+Pb collisions is dominated by
the $\gamma g$ contribution.  The maximum value of $S^i_A$ in the
antishadowing region is $\approx 1.07$ for
EKS98 and $\approx 1.1$ for FGS, reflecting the high $\mu^2$ behavior of the
shadowing parameterizations.  

The EKS98 ratios for the produced quarks and
antiquarks in Fig.~\ref{jetdir}(c)
follow $R(p_T)$ for the total rather closely
over all $p_T$.  The quark and antiquark ratios 
are slightly above the total at low $p_T$ due to the
small $\gamma q$ contribution.  They continue to follow the total at high
$p_T$ since all the EKS98 ratios exhibit similar behavior at large $x$.  The
produced gluon ratio follows the quark ratios for
$p_T > 200$ GeV.  The
large $p_T$ contribution arises from the valence quarks.  Some antishadowing
is observed at low $p_T$ due to the valence quark contribution.  The total
ratios for the lighter ions are closer to unity for all $p_T$ due to their
smaller $A$.  

The results are similar for FGS, shown in Fig.~\ref{jetdir}(d), 
but there are some subtle
differences.  The ratio $R(p_T)$ for produced gluons, arising 
from the $\gamma q$ contribution, exhibits a larger antishadowing effect on 
$R(p_T)$ because $S^{\overline q}_A$ 
is higher for this parameterization, 
see Fig.~\ref{shadcomp}.   The FGS antiquark shadowing
ratio goes to unity for $x>0.2$, flattening $R(p_T)$ for
antiquarks (dotted curve) due to the contribution from $\gamma \overline q
\rightarrow g \overline q$.  The FGS valence quark ratio is taken from
EKS98, resulting in the similarity of $R(p_T)$ in Figs.~\ref{jetdir}(c) and (d)
at high $p_T$.

Recall that some care must be taken when
applying these parameterizations to high $p_T$ since the upper limit of their
fit range is 100 GeV.  While no extraordinary effects are seen in their 
behavior beyond this scale, 
the results should be taken as indicative only.  Finally, we
remark that we have only considered the range $|y_1|\leq 1$.  Including
contributions from all rapidities would increase the effect of shadowing since
smaller $x$ values can be reached when $|y_1|$ is large.  

\begin{figure}[htbp]
\setlength{\epsfxsize=0.95\textwidth}
\setlength{\epsfysize=0.55\textheight}
\centerline{\epsffile{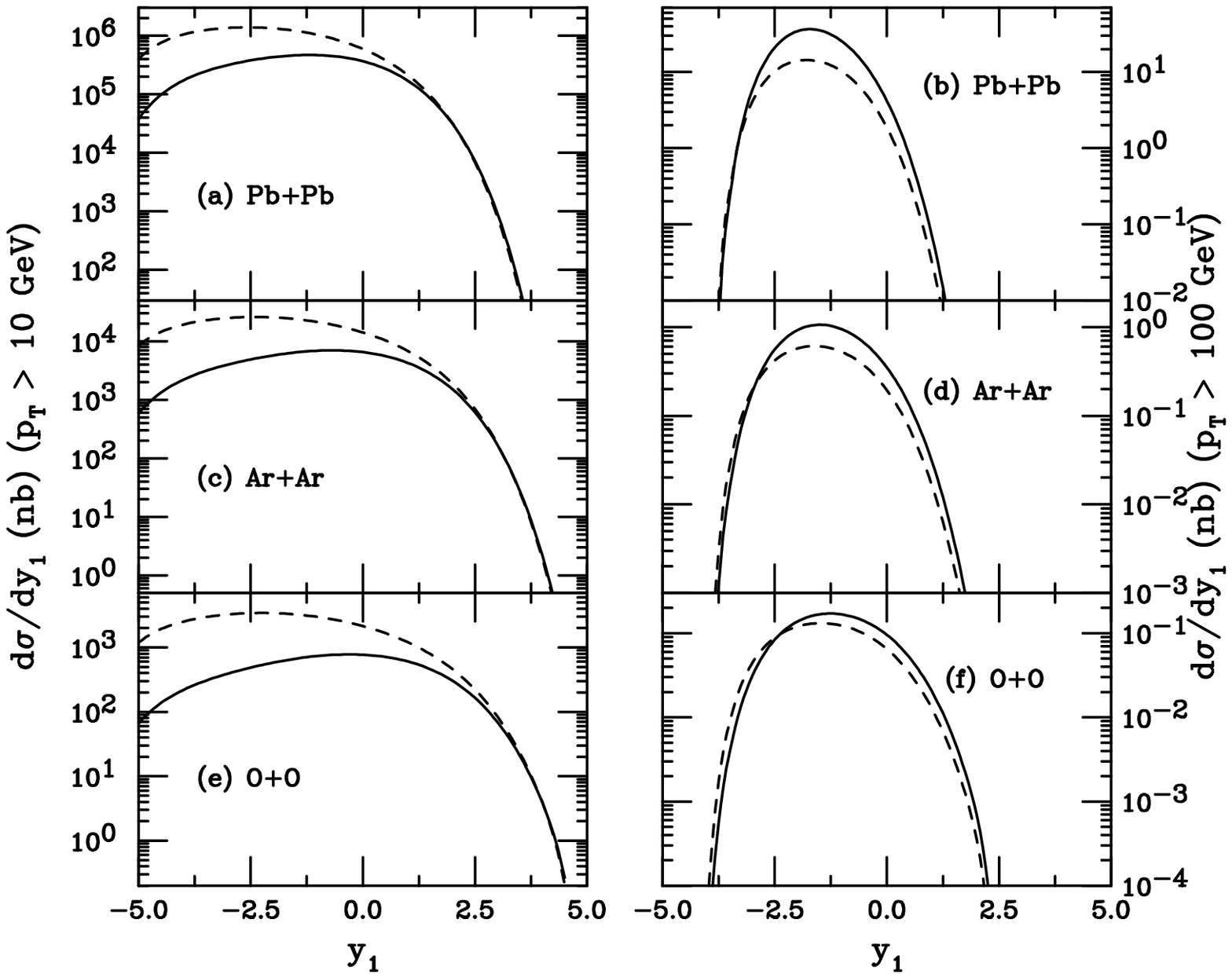}}
\caption[] {We compare the rapidity distributions of direct and resolved dijet
production without shadowing
in peripheral collisions. The left-hand side shows the results
for $p_T > 10$ GeV for (a) Pb+Pb, (c) Ar+Ar and (e) O+O collisions
while the right-hand side is for $p_T > 100$ GeV for (b) Pb+Pb, (d) Ar+Ar
and (f) O+O collisions.  The solid curves are 
the direct results while the dashed curves 
show the resolved results. The photon is coming from the left.}
\label{jet_dir_rap}
\end{figure}

In Fig.~\ref{jet_dir_rap}, we present the rapidity distributions with two 
different values of the minimum $p_T$, $p_T > 10$ GeV on the 
left-hand side and 100 GeV on the right-hand side.  The results, given by the
solid curves, are shown without nuclear shadowing effects.  Note that, in
this case, the photon comes from the left.  There is a symmetric case where
the photon comes from the right, the factor of two on the $p_T$ distribution
in Eq.~(\ref{jetdireq}).  In this case, the $y_1$ distribution is reflected
around $y_1 = 0$.  With the 10 GeV cut, the distributions are rather broad
at negative $y_1$ where the photon has small momentum and, hence, large flux.
At large $y_1>0$, corresponding to small $x$ for the nucleon momentum fractions
and high photon momentum, the distributions fall rapidly since at high photon
momenta, the photon flux is cut off as $k \rightarrow k_{\rm max}$.  The
distributions with the 100 GeV cutoff are narrower because the edge of phase 
space is reached at lower values of $y_1$.  The rapidity distributions are 
broader in general for the lighter systems due to the higher $\sqrt{S_{NN}}$.

\begin{figure}[htbp]
\setlength{\epsfxsize=0.95\textwidth}
\setlength{\epsfysize=0.55\textheight}
\centerline{\epsffile{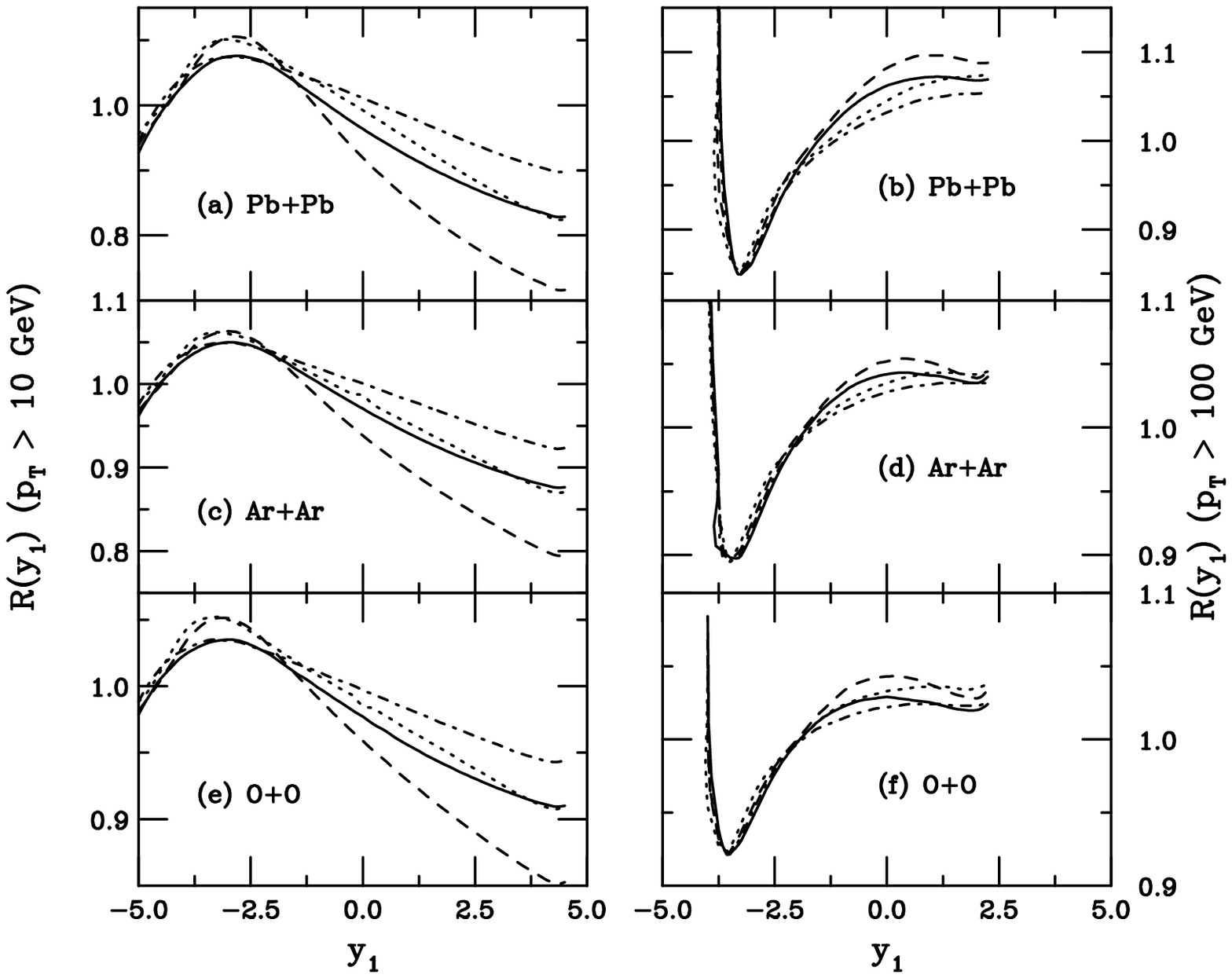}}
\caption[] {We compare shadowing ratios in direct and resolved dijet
production in peripheral collisions. The left-hand side shows the results
for $p_T > 10$ GeV for (a) Pb+Pb, (c) Ar+Ar and (e) O+O collisions
while the right-hand side is for $p_T > 100$ GeV for (b) Pb+Pb, (d) Ar+Ar
and (f) O+O collisions.  The solid and dashed curves give 
the direct ratios for the EKS98 and FGS parameterizations
respectively.  The dot-dashed and dotted curves 
show the resolved ratios for the EKS98 and FGS
parameterizations respectively.  The photon comes from the left.  Note the
difference in the $y$-axis scales here.}
\label{jet_dir_shad}
\end{figure}

Figure~\ref{jet_dir_shad} gives the ratio $R(y_1) = (d\sigma[S_A^i]/dy_1)/
(d\sigma[S_A^i = 1]/dy_1)$ for the two $p_T$ cuts.  The ratios reflect the 
direction of the photon, showing an antishadowing peak at $y_1 \sim -3$, an
EMC region at $y_1 < -4$ and a shadowing region for $y_1 > -0.5$ for $p_T >
10$ GeV, the left-hand side of Fig.~\ref{jet_dir_shad}.  The shadowing
effect is not large, $(20-25)$\% at $y_1
\sim 4$ for Pb+Pb collisions and decreasing with $A$.  The antishadowing peak
is higher for FGS while its shadowing effect is larger at positive $y_1$, as
also noted in the $p_T$-dependent ratios in Fig.~\ref{jetdir}.  A comparison
of the average effect around $|y_1| \leq 1$ with the $p_T$ ratios 
shown in Fig.~\ref{jetdir}, are in good agreement.  Even though $x_2$
is smaller for the lighter systems, the shadowing effect is also reduced.
Since shadowing also decreases with $\mu^2$, the effect is even smaller for
$p_T > 100$ GeV, only $\sim 5$\% at $y_1 \geq 0$, as shown on the right-hand 
side of Fig.~\ref{jet_dir_shad}.  Here
the rise at $y_1 < -3.5$ is the Fermi motion as $x_2 \rightarrow 1$.  At 
$y_1 > -1.2$, the antishadowing region is reached.  

The total dijet photoproduction cross sections without shadowing and the
corresponding rates in a one month, $10^6$ s, LHC run are shown in
Table~\ref{jetdirrates}.
The rates are based on the luminosities of Table~\ref{gamfacs}.  All results
are given in the rapidity interval $|y_1| \leq 1$.  Extended rapidity
coverage, corresponding to  {\it e.g.} $|y_1| \leq 2.4$ for the CMS barrel and 
endcap systems, could increase the rates by a factor of $\approx 2$.  (The
increase in rate with rapidity acceptance is not linear in
$|y_1|$ because the rapidity distributions are asymmetric around $y_1 = 0$
and increasing $p_T$ narrows the rapidity distribution.  The effect of changing
the $y_1$ cut is closer to linear at low $p_T$ and larger at high $p_T$
because the peak is at $y_1 < -1$ for large $p_T$, as seen in the $y_1$ 
distributions on the right-hand side of Fig.~\ref{jet_dir_rap}.) 
Note that by $p_T \approx 
100$ GeV, only a few events are expected per month.  However, while high $p_T$
jets are rare, they should be easily observable in a clean environment like 
photoproduction.

There is a difference of $\approx 500$ in the Pb+Pb and O+O cross sections at
$p_T \approx 11$ GeV, decreasing to less than a factor of four at $\approx 400$
GeV.  The difference decreases with $p_T$ due to the larger phase space
available at high $p_T$ for the higher $\sqrt{S_{NN}}$ systems.
Note that the rates are nearly the same for all systems because even though $A$
is larger for Pb+Pb, higher luminosities and higher $\sqrt{S_{NN}}$ compensates
for the lower $A$ in lighter systems. 

We point out that the jet hadroproduction rates are much higher because they
are multiplied by $\approx A^2$ for hard processes which increase with the
number of binary collisions in $AA$ collisions.  (The relation is not exact
due to shadowing.)  The integration over all impact parameters leads
to $\approx A^2$ scaling while there is only a factor of $A$ in the dijet
photoproduction rate since the photon flux is already integrated over
impact parameter for $b > 2R_A$.  This, combined with the lower effective
energy
and fewer channels, considerably reduces the photoproduction rates relative
to hadroproduction. 

\subsection{Direct leading hadron production}

We now turn to a description of the final-state particle production in the
hadronization of jets.  The particle with the highest $p_T$ is called the
``leading'' particle. 
The corresponding leading particle $p_T$ distribution
is \cite{Field:uq}
\begin{eqnarray}
\label{haddireq}
\frac{d\sigma_{\gamma A \rightarrow hX}^{\rm dir}}{dp_T d^2b} & = &
4 p_T \int dz \int_{\theta_{\rm min}}^{\theta_{\rm max}} 
\frac{d\theta_{\rm cm}}{\sin
\theta_{\rm cm}} \int dk \frac{d^3N_\gamma}{dk d^2b}
\int \frac{dx_2}{x_2} \\ &  & \mbox{} \times \bigg[ \sum_{i,l =
q, \overline q, g} F_i^A(x_2,\mu^2, \vec b, z) 
\frac{d\sigma_{\gamma i \rightarrow lX'}}{dt} 
\frac{D_{h/l}(z_c,\mu^2)}{z_c} \bigg] \, \,  \nonumber
\end{eqnarray}
where the $X$ on the left-hand side includes all final-state hadrons in
addition to $h$ but
$X'$ on the right-hand side denotes the unobserved final-state parton.
The subprocess cross sections, $d\sigma/dt$, 
are related to 
$s^2 d\sigma/dt du$ in Eq.~(\ref{jetdireq})
through the momentum-conserving 
delta function $\delta(s + t + u)$ and division by $s^2$.
The integrals over rapidity have been replaced by an integral over
center-of-mass scattering angle, $\theta_{\rm min} \leq 
\theta_{\rm cm} \leq \theta_{\rm max}$, corresponding to a given
rapidity cut.  Here
$\theta_{\rm min} =0$ and $\theta_{\rm max} = \pi$ covers the full rapidity
range while $\theta_{\rm min} = \pi/4$ and $\theta_{\rm max} = 3\pi/4$ roughly
corresponds to $|y_1| \leq 1$. 
The fraction of the final
hadron momentum relative to that of the produced parton, $z_c$, appears 
in the fragmentation function, $D_{h/l}(z_c,\mu^2)$, the probability to 
produce hadron
$h$ from parton $l$.  The fragmentation functions are assumed to be universal,
independent of the initial state.

The produced partons are fragmented into charged pions, charged kaons and
protons/antiprotons using LO fragmentation functions fit to $e^+ e^-$ data
\cite{Kniehl:2000fe}.  The final-state hadrons are assumed to be produced
pairwise so that $\pi \equiv (\pi^+ + \pi^-)/2$, $K \equiv (K^+ +
K^-)/2$, and $p \equiv (p + \overline p)/2$.  The equality of $p$ and
$\overline p$ production obviously does not describe low energy hadroproduction
well.  As energy increases, this approximation may become more
reasonable.  The produced hadrons follow the parent parton
direction.  
We have used the LO KKP fragmentation functions \cite{Kniehl:2000fe}.  
The KKP scale evolution is modeled using $e^+e^-$ data at
several different energies and
compared to $p \overline p$, $\gamma p$ and $\gamma\gamma$ data.
After some slight scale modifications \cite{Kniehl:2000hk} all the $h^-$ data
could be fit.  However, there are significant
uncertainties in fragmentation when the leading hadron takes most of
the parton momentum \cite{Zhang:2002py}, as is the case here.

We assume the same scale in the parton densities 
and the fragmentation functions, $\mu^2 =
p_T^2$.  A larger scale, $p_T^2/z_c^2$, has sometimes been used in the parton
densities.  At high
$p_T$, where $z_c$ is large, the difference in the results for the two scales
is small.  We have not included any intrinsic transverse momentum broadening
in our calculations \cite{Gyulassy:2001nm,Vitev:2002vr}.  
This ``Cronin'' effect can be important when $p_T$ is small
but becomes negligible for transverse momenta larger than a few GeV.

The largest contribution to the total final-state charged particle production
is charged pions, followed by kaons.  The proton contribution is 
smallest even though at RHIC the $p/\pi$ ratio approaches unity in central and
0.4 in peripheral Au+Au collisions \cite{Sakaguchi:2002bm} for $p_T \sim
2$ GeV.  This discrepancy is due to the poor knowledge of the fragmentation 
functions at large $z_c$, particularly for baryons.
Hopefully by the time the LHC begins operation, updated
fragmentation functions incorporating $pp$ data from RHIC will be available,
allowing better estimates of leading particle production at higher $z_c$.

\begin{figure}[htbp]
\setlength{\epsfxsize=0.95\textwidth}
\setlength{\epsfysize=0.45\textheight}
\centerline{\epsffile{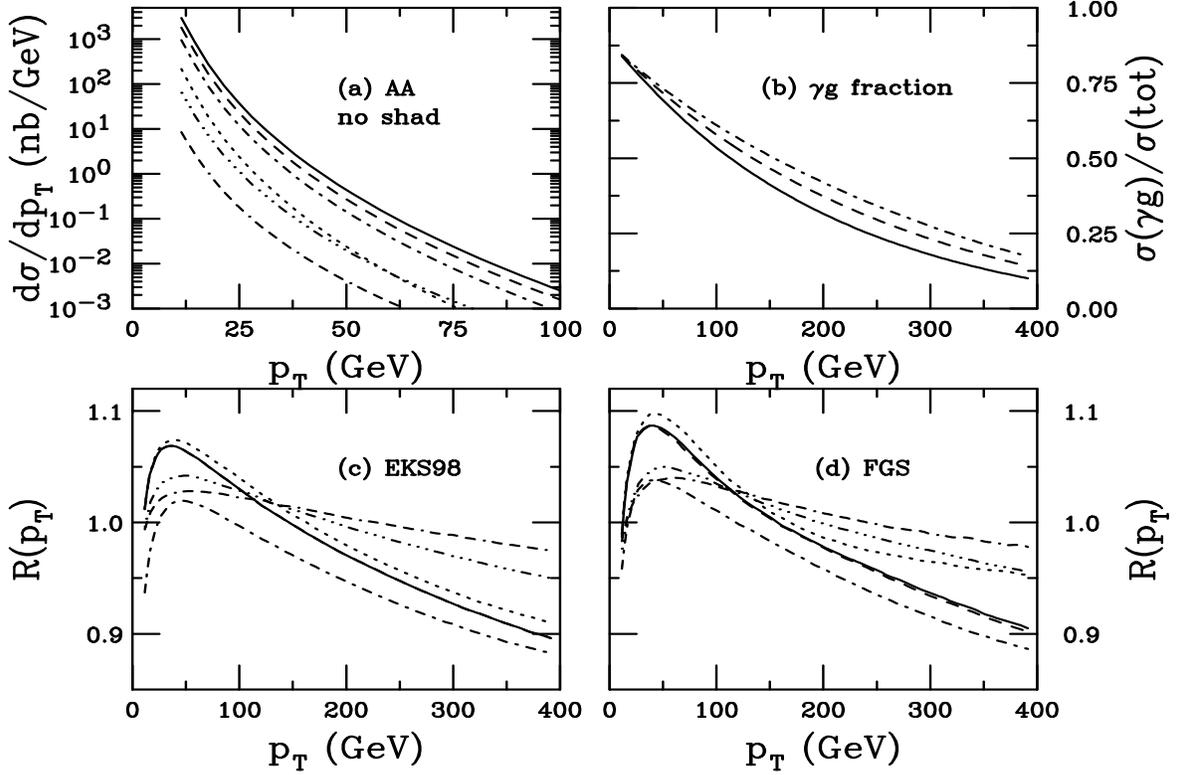}}
\caption[] {Direct photoproduction of leading hadrons 
in peripheral collisions. (a)
The $p_T$ distributions for $|y_1| \leq 1$ are shown for $AA$ 
collisions.  The solid curve is the total for Pb+Pb while the produced 
pions (dashed), kaons (dot-dashed) and protons (dotted) are shown
separately.  The total production for Ar+Ar (dot-dot-dot-dashed) and O+O
(dot-dash-dash-dashed) are also shown. (b) The fraction of gluon-initiated
hadrons as a function of $p_T$.  The curves are the same as in (a). 
(c) The EKS98 shadowing ratios for produced pions.
The solid curve is the total for Pb+Pb while the ratios for pions produced by
quarks (dashed), antiquarks (dotted) and gluons (dot-dashed) are shown
separately.  The total ratios for Ar+Ar (dot-dot-dot-dashed) and O+O
(dot-dash-dash-dashed) are also shown. (d) The same as (c) for FGS.}
\label{jethaddir}
\end{figure}

The corresponding hadron distributions from direct jet photoproduction
are shown in Fig.~\ref{jethaddir}(a) for $AA$ collisions.
Note that the leading hadron cross sections are lower than the partonic
jet cross sections, compare 
Figs.~\ref{jetdir}(a) and \ref{jethaddir}(a).  Several 
factors can account for this.  The maximum $\sqrt{S_{\gamma N}}$ is a factor 
of five or more less than $\sqrt{S_{NN}}$ for $AA$ collisions.  The reduced 
number of processes available for direct dijet 
photoproduction is a significant contribution to the decrease.
Note also that the $p_T$ distribution is steeper for leading hadrons than for
the jets, as may be expected since the effective $p_T$ of the hadron
is higher than than of the produced parton.

The average $z_c$ for direct photoproduction of high $p_T$ particles 
is $\approx 0.4$
for particles with $p_T \approx 10$ GeV, increasing to $\langle z_c 
\rangle > 0.45-0.55$ for $p_T > 100$ GeV.  The lower $z_c$ values correspond to
lighter ion collisions.  In this $z_c$ region, the fragmentation
functions are not very well known.  As pointed 
out in Ref.~\cite{Zhang:2002py},
a small change in the fragmentation function fits can produce significant
changes at large $z_c$.  This region is not well constrained by
the $e^+ e^-$ data used in the fits.

The effect of fragmentation on the production channels is shown in
Fig.~\ref{jethaddir}(b) where we present the fraction of leading hadron 
production
from the $\gamma g$ channel for all charged hadrons.
The ratios are rather similar to those of the partonic jets although the
$\gamma g$ fraction is somewhat smaller due to the larger average $x$ of hadron
production with respect to jets, as we discuss later.

The shadowing ratios for charged pions produced in Pb+Pb collisions by quarks, 
antiquarks, gluons and the total from all partons,
are shown for the EKS98 and FGS parameterizations
in Fig.~\ref{jethaddir}(c) and (d).  The ratios for pion production in Ar+Ar
and O+O collisions are also shown.  The high $p_T$ FGS antiquark
ratios flatten out relative to the EKS98 ratio
because the $\gamma \overline q$ channel dominates gluon production at
high $p_T$.  The flattening begins at lower $p_T$ here because the $x$
for hadron production is higher than that for the jets.
The ratio of pions arising from produced gluons follows the valence ratio, as
expected. 
The ratios decrease with increasing $p_T$ due to 
the EMC effect for $x > 0.2$ when $p_T > 100$ GeV.  

\begin{figure}[htbp]
\setlength{\epsfxsize=0.95\textwidth}
\setlength{\epsfysize=0.45\textheight}
\centerline{\epsffile{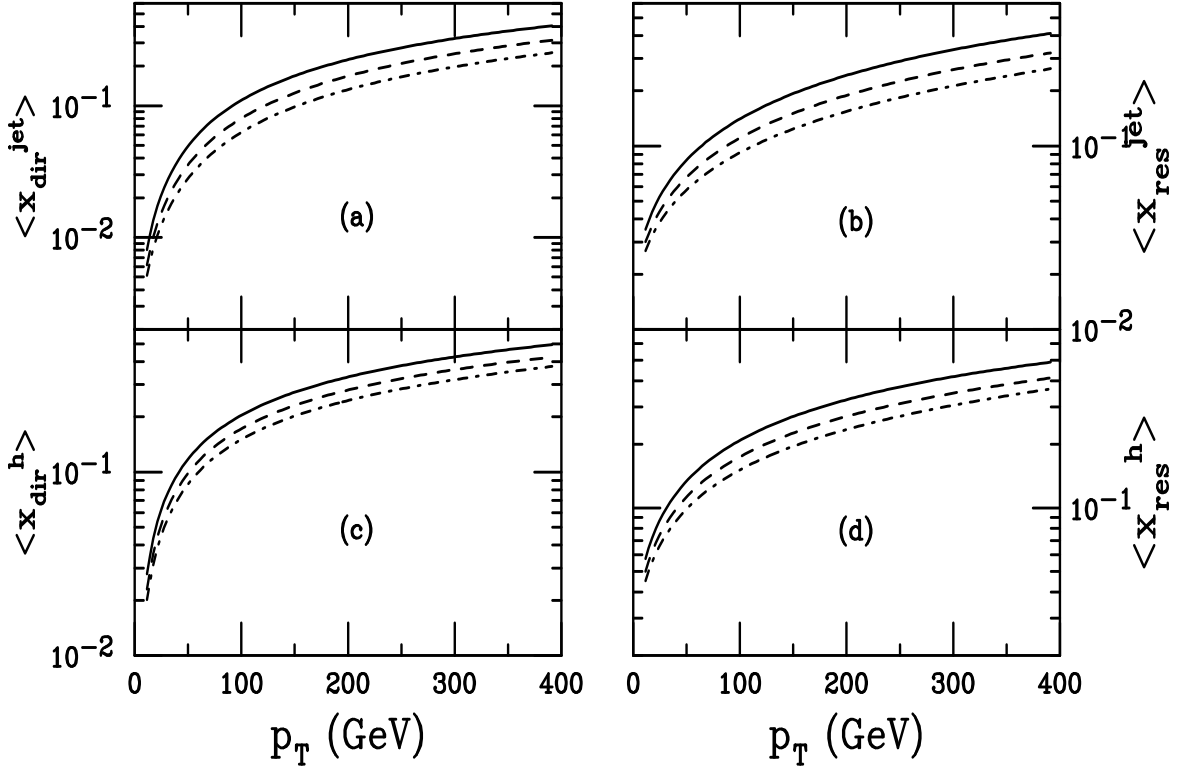}}
\caption[]{The average value of the nucleon parton momentum fraction $x$ as a
function of transverse momentum.  Results are given for (a) direct and (b) 
resolved gluon jet production and for (c) direct and (d) resolved pion 
production by gluons.  The results are given for
O+O (dot-dashed), Ar+Ar (dashed) and Pb+Pb (solid) interactions.}
\label{avexjet}
\end{figure}

We now discuss the relative values of the nucleon momentum fraction, $x$ for
parton and hadron production.  
On the left-hand side of Fig.~\ref{avexjet}, we compare the average $x$ values
for produced gluon jets (upper plot) and for pions produced by these gluons 
(lower plot).
We have chosen to compute the results for produced gluons alone to better
compare with resolved jet photoproduction, discussed next.  
Since we are interested in produced gluons, we only
consider the QCD Compton contribution, $\gamma q \rightarrow gq$.  
This channel is biased toward larger
momentum fractions than $\gamma g \rightarrow q \overline q$
since the gluon distribution is largest at 
small $x$ while the valence quark
distribution in the proton is peaked at $x \sim 0.2$. The average $x$ for 
a gluon jet is $\sim 0.005-0.008$ at $p_T\approx 10$ GeV, increasing to
$\sim 0.03-0.05$ at 50 GeV.
The smallest $x$ is from the highest energy O+O collisions.  The average $x$
increases with $p_T$, to $\sim 0.25-0.4$ at $p_T \sim 400$ GeV.
When final state pions are considered, in the lower left-hand plot, at low
$p_T$, the average $x$ is larger than for gluon jets.  At $p_T \approx 10$ GeV,
$\langle x \rangle \approx 0.02-0.03$ while at 50 GeV, 
$\langle x \rangle \approx 0.09 - 0.12$.  At high $p_T$, however, the
average $x$ becomes similar for jet and hadron production as
$\langle z_c \rangle$ increases to $\approx 0.6-0.7$ at $p_T \sim 400$ GeV.

\subsection{Resolved dijet production}

We now turn to resolved production.  
The hadronic reaction, $\gamma N
\rightarrow {\rm jet} \, + {\rm jet}\, + X$, is unchanged, but in this case, 
prior to the
interaction with the nucleon, the photon splits into a color singlet
state of $q \overline q$ pairs and gluons.  
On the parton level, the resolved LO reactions are {\it e.g.} $g(xk) +
g(x_2 P_2) \rightarrow g(p_1) + g(p_2)$ where $x$ is the
fraction of the photon momentum carried by the parton.  The LO processes for
resolved photoproduction, illustrated in Fig.~\ref{dijetdia}(c)-(h), 
are the same as those for LO $2 \rightarrow 2$
hadroproduction except that one parton source is a photon
rather than a nucleon.
The resolved jet photoproduction cross
section for partons of flavor $f$ in the
subprocess $ij\rightarrow kl$ in $AB$ collisions is, modified from 
Refs.~\cite{Emel'yanov:1999bn,Eskola:1988yh,Eskola:1996ce},
\begin{eqnarray} S_{NN}^2
\frac{d\sigma^{\rm res}_{\gamma A \rightarrow {\rm jet} \, + \, {\rm 
jet}}}{dT dU d^2b} & = & 2 \int dz \int_{k_{\rm min}}^\infty 
\frac{dk}{k} {d^3N_\gamma\over dkd^2b} \int_{k_{\rm min}/k}^1 \frac{dx}{x}
\int_{x_{2_{\rm min}}}^1 \frac{dx_2}{x_2} \nonumber \\
&  & \mbox{} \times \sum_{{ij=}\atop{\langle kl \rangle}} \left\{
F_i^\gamma (x,\mu^2) F_j^A(x_2,\mu^2,\vec b,z) + F_j^\gamma (x,\mu^2)
F_i^A(x_2,\mu^2,\vec b,z) \right\} \nonumber \\ 
&  & \mbox{} \times  \frac{1}{1 + \delta_{kl}}
\left[\delta_{fk} \hat{s}^2\frac{d\sigma^{ij\rightarrow kl}}{d\hat t d\hat u}
(\hat t, \hat u) 
+ \delta_{fl} \hat{s}^2 \frac{d\sigma^{ij\rightarrow kl}}{ d\hat t d\hat u}
(\hat u, \hat t)
\right] \, \, 
\label{sigjetres} 
\end{eqnarray}
where $\hat{s} = (xk + x_2P_2)^2$, $\hat{t} = (xk - p_1)^2$,
and $\hat{u} = (x_2P_2 - p_1)^2$.  
The $2 \rightarrow 2$ minijet subprocess cross sections, $d\sigma/d\hat t$, 
given in the review article by Owens \cite{Owens:1986mp}, are related to 
$d\sigma/d\hat t d\hat u$ through the momentum-conserving 
delta function $\delta(\hat s + \hat t 
+ \hat u)$.
The sum over initial states includes all combinations of two parton
species with three flavors while the final state includes all pairs
without a mutual exchange and four flavors (including charm).  The
factor $1/(1 + \delta_{kl})$ accounts for identical particles in
the final state.

\begin{figure}[htbp]
\setlength{\epsfxsize=0.95\textwidth}
\setlength{\epsfysize=0.45\textheight}
\centerline{\epsffile{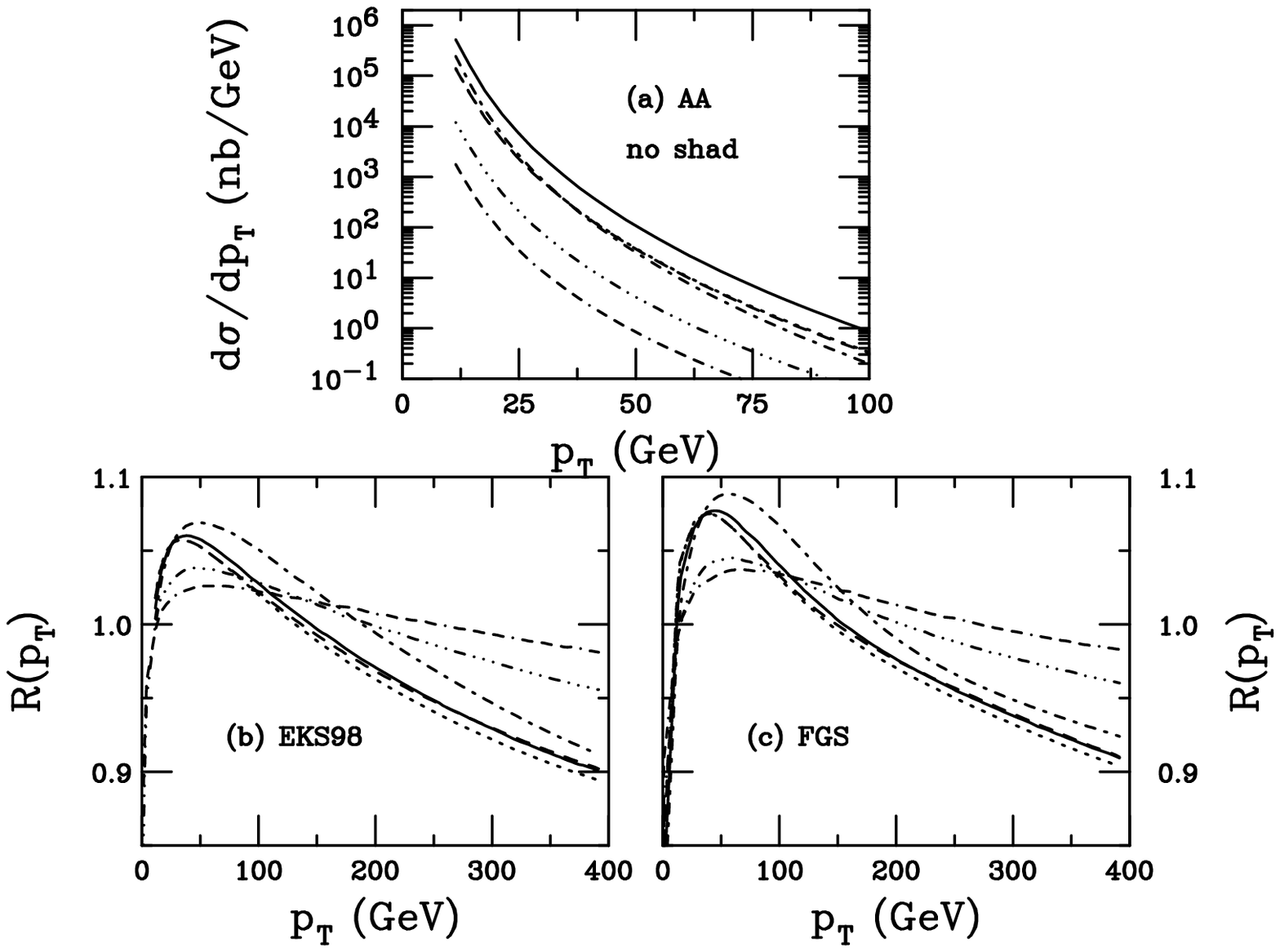}}
\caption[] {Resolved dijet photoproduction in peripheral $AA$ collisions. (a)
The Pb+Pb jet $p_T$ distributions with $|y_1| \leq 1$ are shown for quarks 
(dashed),
antiquarks (dotted), gluons (dot-dashed) and the total (solid).  We also show
the total jet $p_T$ distributions in Ar+Ar (dot-dot-dot-dashed) and 
O+O (dash-dash-dash-dotted) collisions. (b) The relative EKS98 shadowing 
contributions 
from quarks (dashed), antiquarks (dotted) and gluons (dot-dashed) as well
as the total (solid) are shown for Pb+Pb collisions.  The totals are also shown
for Ar+Ar (dot-dot-dot-dashed) and O+O (dash-dash-dash-dotted) interactions.
(c) The same as (b) for FGS.}
\label{pjetres}
\end{figure}

The resolved jet results, shown in Fig.~\ref{pjetres},
are independent of the photon parton densities for $p_T>10$ GeV.  
Along with the total quark, antiquark and gluon cross sections 
in Pb+Pb collisions, we also show the
individual partonic contributions to the jet $p_T$ distributions.  The 
produced gluon contribution dominates for $p_T < 25$ GeV but, by 50 GeV, 
quark and antiquark production becomes larger due to the increase of the $qg
\rightarrow qg$ channel relative to the $gg \rightarrow gg$ channel.  
We also show the total $p_T$ distributions
for Ar+Ar and O+O collisions.
For lighter nuclei, the crossover between gluon and quark/antiquark dominance
occurs at higher $p_T$ due to the higher collision energy.

Note that the resolved dijet photoproduction contribution is larger than the
direct by a factor of two to three for $p_T<50$ GeV, despite the
lower effective center of mass energy of resolved production.  
Resolved production opens up many more channels through the parton components
of the photon.  Indeed, now all the $2 \rightarrow 2$ channels for LO jet
hadroproduction are available.  In addition, the quark and antiquark
distributions in the photon are the same.  These distributions are large 
at high momentum fractions, higher 
than the quark and antiquark distributions in the proton.  Thus
the quark and antiquark channels are enhanced relative to hadroproduction.
The largest difference between the quark and antiquark production rates 
is due to the difference between the valence and sea quark distributions in the
nucleus.  Where the valence and sea 
quark contributions are similar, as for $|y_1|\leq 1$, the difference is rather
small.  If all rapidities were included, the relative quark and antiquark
rates could differ more.

The direct and resolved rapidity distributions are compared in
Fig.~\ref{jet_dir_rap} for the two $p_T$ cuts, 10 and 100 GeV.  While the
$|y_1|\leq 1$ resolved contribution is a factor of two to three
larger than the direct at $p_T < 50$ GeV, a comparison of the $y_1$ 
distributions over all rapidities shows that the resolved contribution can be
considerably larger, a factor of $\sim 5-10$ at $y_1 < -3.5$ for $p_T
> 10$ GeV.  At $p_T > 100$ GeV, the resolved contribution is still equivalent
to or slightly larger than the direct at $y_1 < -3$ but drops below at
larger rapidities.  Thus, going to higher $p_T$ can separate direct from
resolved production, especially at forward rapidities.  
Recall that the produced gluons 
dominate resolved production at $p_T < 25$ GeV while they are only a small
contribution to direct production.  The largest gluon production channels are
typically $gg \rightarrow gg$ and $gq \rightarrow gq$.  As $y_1$ becomes large
and negative, the photon $x$ decreases while $x_2$ of the nucleon increases,
leading to the dominance of the $gq$ channel.  The photon gluon distribution
is largest as $x$ decreases.  The valence quark distribution of the proton
is also important at high $p_T$, causing the resolved to direct ratio to
flatten for $y_1 > -2.5$ when $p_T >100$ GeV.

In Fig.~\ref{jet_dir_shad}, we compare the direct and resolved shadowing 
ratios,
$R(y_1)$.  A smaller shadowing effect is observed for the resolved component
due to the larger $x_2$ for resolved production.  The difference
in the direct and resolved shadowing ratios is reduced for larger $p_T$.

\begin{figure}[htbp]
\setlength{\epsfxsize=0.95\textwidth}
\setlength{\epsfysize=0.3\textheight}
\centerline{\epsffile{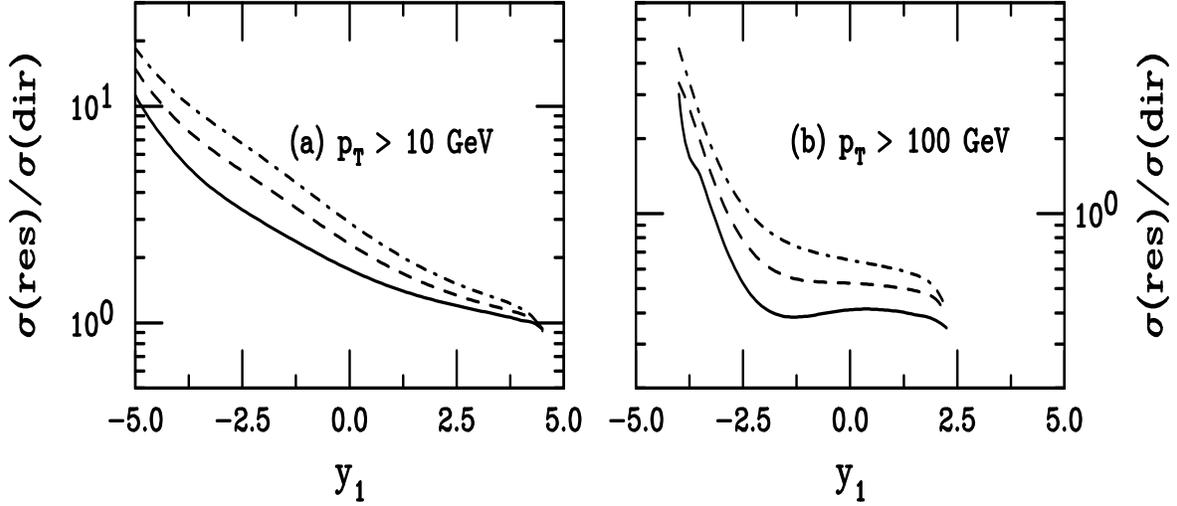}}
\caption[] {We present the resolved/direct dijet production ratios as a 
function of rapidity. In (a) we show the results
for $p_T > 10$ GeV 
while in (b) we show $p_T > 100$ GeV.  
The curves are Pb+Pb (solid), Ar+Ar (dashed) and O+O (dot-dashed).  The photon
comes from the left.}
\label{jet_dir_rat}
\end{figure}

To measure the nuclear parton densities most directly, it is preferable for
direct production to be dominant.  However, Fig.~\ref{jet_dir_rat} shows that a
$p_T$ cut is not very
effective for dijet production, as previously
mentioned, even at forward rapidity.  The resolved to direct production ratios
are all larger than unity for $p_T > 10$ GeV, even for large, positive $y_1$.
While the ratio is less than 1 for $y_1 > -2.5$ and $p_T >
100$ GeV, it is only $\sim 0.5$ for Pb+Pb, increasing to 0.8 for O+O.

Thus, although clean separation is possible at $p_T > 100$ GeV, precision
parton density measurements are not possible at these values of
$p_T$ due to the low
rate.  Other means of separation must then be found.  Resolved processes will
not be as clean as direct in the direction of the photon due to the breakup of
the partonic state of the photon.  The multiplicity in the photon fragmentation
region will be higher than in direct production where the nucleus should remain
intact.  A cut on multiplicity in the photon direction may help separate the
two so that, although there should be a rapidity gap for both direct and 
resolved photoproduction, the gap may be less prominent for resolved 
production.

The total resolved dijet photoproduction cross sections without shadowing and
the corresponding monthly ($10^6$ s) LHC rates are given in
Table~\ref{jetresrates}.  At low $p_T$, the cross sections and rates are a
factor of $2-3$ higher than for direct dijet production.  The largest 
increase is for
the lightest nuclei since the lowest $x$ values are probed.  However, with
increasing $p_T$, the phase space is reduced.  The average
photon momentum is increased and, at large photon momentum, the flux drops
faster.
The average momentum fractions probed by the nuclear parton densities grows
large and only valence quarks contribute.  The lower effective energy of
resolved relative to direct photoproduction reduces the high $p_T$ phase space
for resolved production.  Thus, at the highest $p_T$, the rate is reduced
relative to direct by a factor of $4-9$ with the smallest decrease for the
lightest system due to the higher effective $\sqrt{S_{NN}}$.
Since resolved production has a narrower rapidity distribution than direct
production, increasing the rapidity coverage would not increase the rate
as much as for direct photoproduction.

We show the individual partonic shadowing ratios for Pb+Pb collisions with
the EKS98 parameterization in
Fig.~\ref{pjetres}(b).  The quark and antiquark shadowing ratios are 
very similar although the quark ratio becomes
larger for higher $p_T$ (higher $x$) values of $x$ due to the valence 
distribution.  Now the gluon ratio shows larger antishadowing since gluon 
production is dominantly 
through the $gg$ and $qg$ channels rather than $\gamma q$ in
direct production, compare Fig.~\ref{jetdir}.  The FGS parameterization
gives similar results, Fig.~\ref{pjetres}(c).  However, since the small
FGS gluon antishadowing is stronger, $R(p_T)$ is larger for $p_T < 150$ GeV.

\subsection{Resolved leading particle production}

\begin{figure}[htbp]
\setlength{\epsfxsize=0.95\textwidth}
\setlength{\epsfysize=0.45\textheight}
\centerline{\epsffile{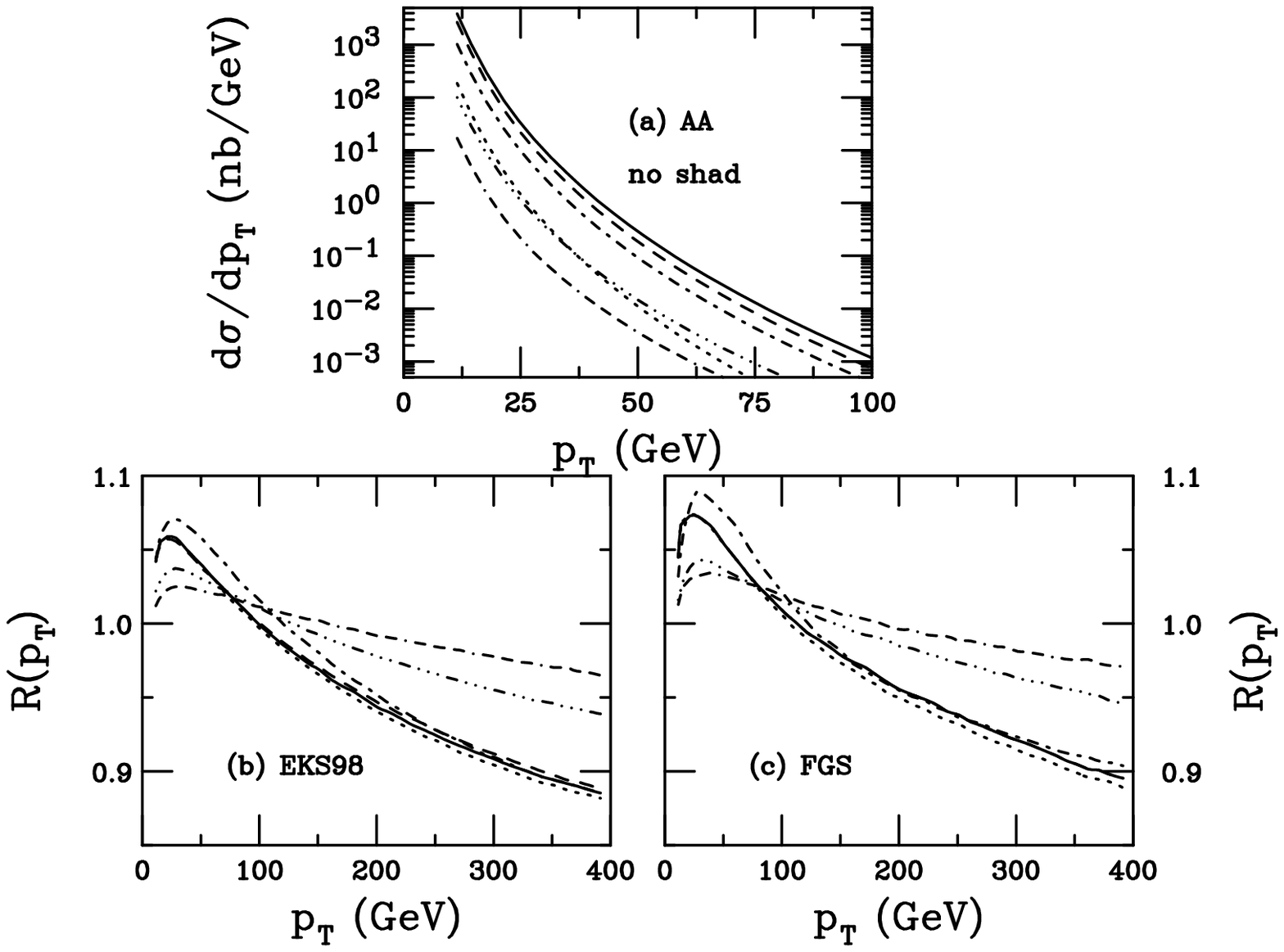}}
\caption[] {Resolved leading hadrons from dijet
photoproduction in peripheral collisions.
(a) The $p_T$ distributions for $|y_1| \leq 1$ are shown for $AA$
collisions.  The Pb+Pb results are shown for charged pions (dashed), kaons 
(dot-dashed), protons (dotted) and the sum of all charged hadrons (solid).   
The charged hadron $p_T$ distributions are also shown for
Ar+Ar (dot-dot-dot-dashed) and O+O (dot-dash-dash-dashed) collisions.
(b) The EKS98 shadowing ratios for produced pions.  For Pb+Pb collisions,
we show the ratios for pions produced by quarks (dashed), antiquarks 
(dotted), gluons (dot-dashed) and the total (solid) separately.  The ratios 
for pions produced by all partons are also shown for Ar+Ar (dot-dot-dot-dashed)
and O+O (dot-dash-dash-dashed) collisions. (c) The same as (b) for FGS.
}
\label{jethadres}
\end{figure}

We now turn to leading particles from resolved jet photoproduction.
The leading particle $p_T$ distributions are
\begin{eqnarray}
\frac{d\sigma^{\rm res}_{\gamma A \rightarrow hX}}{dp_T d^2b} & = & 4p_T \int d
z  
\int_{\theta_{\rm min}}^{\theta_{\rm max}}
\frac{d\theta_{\rm cm}}{\sin \theta_{\rm cm}}
\int_{k_{\rm min}}^\infty 
\frac{dk}{k} {d^3N_\gamma\over dkd^2b} \int_{k_{\rm min}/k}^1 \frac{dx}{x}
\int_{x_{2_{\rm min}}}^1 \frac{dx_2}{x_2} \nonumber \\
&  & \mbox{} \times \sum_{{ij=}\atop{\langle kl \rangle}} \left\{
F_i^\gamma (x,\mu^2) F_j^A(x_2,\mu^2,\vec b,z) + F_j^\gamma (x,\mu^2)
F_i^A(x_2,\mu^2,\vec b,z) \right\} \nonumber \\ 
&  & \mbox{} \times  \frac{1}{1 + \delta_{kl}}
\left[\delta_{fk} \frac{d\sigma^{ij\rightarrow kl}}{d\hat t}(\hat t, \hat
u) 
+ \delta_{fl} \frac{d\sigma^{ij\rightarrow kl}}{d\hat t}(\hat u, \hat t)
\right] \frac{D_{h/k}(z_c,\mu^2)}{z_c} \, \, .
\label{sighadres} 
\end{eqnarray}
The subprocess cross sections, $d\sigma/d\hat t$, 
are related to 
$\hat s^2 d\sigma/d\hat t d\hat u$ in Eq.~(\ref{sigjetres})
through the momentum-conserving 
delta function $\delta(\hat s + \hat t + \hat u)$ and division by $\hat s^2$.
The drop in rate between jets and high $p_T$ hadrons is similar to that in
direct photoproduction, as can be seen by comparison of Figs.~\ref{pjetres}
and \ref{jethadres} relative to Figs.~\ref{jetdir} and \ref{jethaddir}.  
Now that gluon fragmentation makes a larger contribution, the relative
pion contribution is larger than in direct photoproduction while
the relative proton contribution is significantly reduced.
The smaller effective center of mass energy for resolved
photoproduction lowers the phase space available for fragmentation. 
Baryon production is then reduced compared to light mesons.

The reduction in phase space for leading hadrons relative to fast
partons can be seen in the comparison of the average $x$ values for resolved 
photoproduction of jets and leading hadrons, shown on the right-hand side of 
Fig.~\ref{avexjet} for gluons and pions from gluons respectively.  At 
$p_T \approx 10$ GeV, 
the average $x$ of the gluon jet is $0.03 - 0.04$, increasing to $0.16 
- 0.24$ at $p_T \approx 200$ GeV, higher than for direct photoproduction, as
expected.  The $x$ values for hadron production are larger still,
$\approx 0.06$ at low $p_T$ while $x \approx 0.23-0.33$ at $p_T \approx 200$
GeV.

The shadowing ratios in Fig.~\ref{jethadres} also reflect the increasing $x$.
Now the antishadowing peak is shifted to $p_T \approx 30$ GeV since 
the average $x$ values are in the EMC region,
even at low $p_T$.  The values of $R(p_T)$ at high $p_T$ are somewhat lower
than for direct production due to the higher $x$.
The average $z_c$ of the fragmentation functions is also somewhat larger for
resolved production, $0.7 - 0.8$ at $p_T \approx 400$ GeV.  

Since the resolved jet cross section is larger
than the direct at low $p_T$, it is more difficult to make clean
measurements of the nuclear gluon distribution unless the two contributions 
can be separated by other methods.  Instead, the
large valence quark contribution at high $p_T$ suggests that jet 
photoproduction can probe the nuclear valence quark distributions
at larger $\mu^2$ than previously possible.  At $p_T> 100$ GeV, more than 
half of direct jet production is through the $\gamma q$ channel.  
Unfortunately, the rates are low here, making high precision 
measurements unlikely.  However, the events should be very clean.

\section{$\gamma+$jet production}

\subsection{Direct $\gamma+$jet production}

\begin{figure}[htbp]
\setlength{\epsfxsize=0.5\textwidth}
\setlength{\epsfysize=0.35\textheight}
\centerline{\epsffile{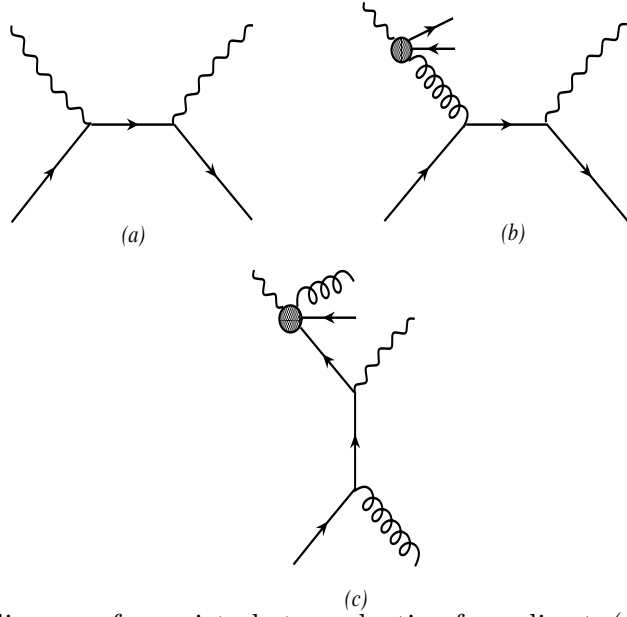}}
\caption[]{Feynman diagrams for $\gamma +$jet photoproduction from direct, (a),
and resolved photons, (b) and (c).  Crossed diagrams are not shown.}
\label{comptdia}
\end{figure}

A clean method of determining the quark distribution in the nucleus
at lower $p_T$ is the
process where a jet is produced opposite a photon in the final state, Compton
scattering in direct production.  The cross sections are reduced relative to
the jet$+$jet process since the coupling is $\alpha^2 e_Q^4$ in
the coupling rather than $\alpha \alpha_s e_Q^2$, 
as in dijet production.  In addition, the 
quark distributions are lower than the gluon, also reducing the rate.
The diagrams for $\gamma+$jet production are shown in Fig.~\ref{comptdia}.

We now discuss the jet and leading particle distributions
for direct and resolved $\gamma +$jet photoproduction.
The hadronic process is $\gamma(k) + N(P_2) \rightarrow
\gamma(p_1) + {\rm jet}(p_2) \, + X$.
The only partonic contribution to the $\gamma+$jet yield in direct 
photoproduction is
$\gamma(k) + q(x_2P_2)  \rightarrow \gamma(p_1) +
q(p_2)$ (or $q \rightarrow \overline q$) where the 
produced quark is massless.  We now have
\begin{eqnarray}
\label{compdireq} S_{NN}^2
\frac{d^2\sigma_{\gamma A \rightarrow \gamma + \, {\rm jet} + X}^{\rm 
dir}}{dT dU d^2b} = 
2 \int dz \int_{k_{\rm min}}^\infty dk \frac{d^3N_\gamma}{dk d^2b}
\int_{x_{2_{\rm min}}}^1 \frac{dx_2}{x_2} \bigg[ \sum_{i=q, \overline q} 
F_i^A(x_2,\mu^2, \vec b, z) s^2 \frac{d^2\sigma_{\gamma i\rightarrow \gamma
i}}{dtdu} \bigg] 
\, 
\end{eqnarray}
where the partonic cross section for the Compton process is
\begin{eqnarray}
\label{sigcompt}
s^2 \frac{d^2\sigma_{\gamma q\rightarrow \gamma q}}{dt du} = - \frac{2}{3} 
\pi \alpha^2 e_Q^4 \bigg[ \frac{s^2 + u^2}{s u}\bigg] \delta(s + t + u) \, \, .
\end{eqnarray}
The extra factor of two on the right-hand side of
Eq.~(\ref{compdireq}) again arises because 
both nuclei can serve as photon sources in $AA$ collisions.  
The kinematics are the same as in jet$+$jet
photoproduction,
described in the previous section.

\begin{figure}[htbp]
\setlength{\epsfxsize=0.95\textwidth}
\setlength{\epsfysize=0.3\textheight}
\centerline{\epsffile{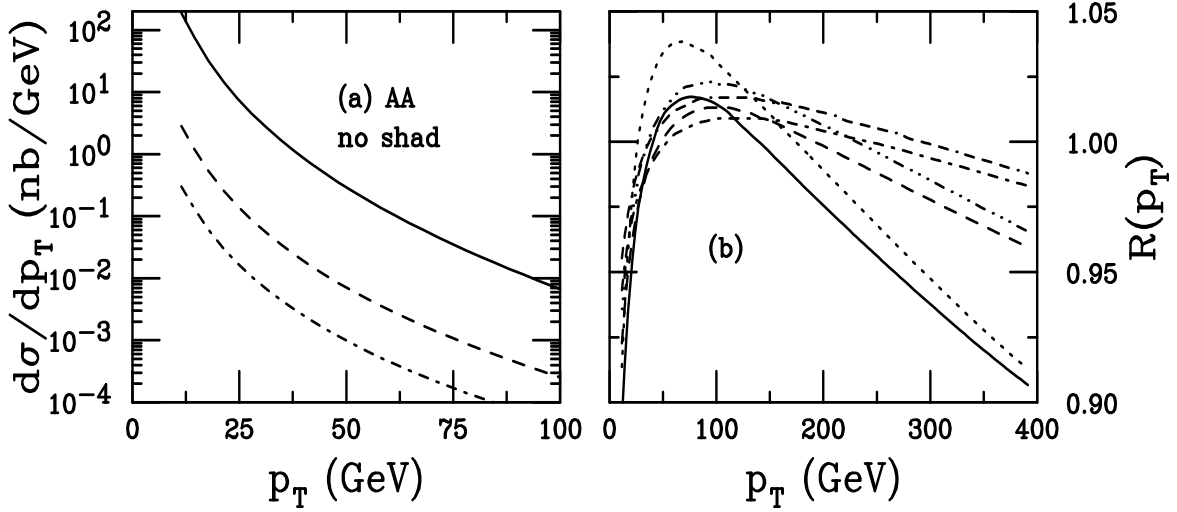}}
\caption[] {Direct $\gamma+$jet photoproduction in peripheral collisions. (a)
The $p_T$ distributions for $|y_1| \leq 1$ are shown for Pb+Pb (solid), Ar+Ar
(dashed) and O+O (dot-dashed) collisions. (b) The EKS98 shadowing ratios are 
shown for Pb+Pb (solid), Ar+Ar (dashed) and O+O
(dot-dashed) while the corresponding FGS ratios are shown for
Pb+Pb (dotted), Ar+Ar
(dot-dot-dot-dashed) and O+O (dot-dash-dash-dashed) collisions.  
The photon comes from the left.
}
\label{compdir}
\end{figure}

The direct $\gamma +$jet photoproduction results are given in 
Fig.~\ref{compdir}  
for $AA$ interactions at the LHC.  We
show the transverse momentum, $p_T$, distributions for all produced 
quarks and antiquarks in Pb+Pb, Ar+Ar and O+O collisions for $|y_1|\leq 1$.

There is a drop of nearly three orders of magnitude between the dijet
cross sections in Fig.~\ref{jetdir} and the $\gamma+$jet cross sections 
in Fig.~\ref{compdir}.  Most of this
difference comes from the relative couplings.  The
rest is due to the reduced number of channels available for direct $\gamma+$jet
production since more than half of all directly produced are gluon-initiated
for $p_T < 100$ GeV, see Fig.~\ref{jetdir}(b).

We have not distinguished between the quark and antiquark initiated jets.  
However, the quark-initiated jet
rate will always be somewhat higher due to the valence contribution. 
When $p_T < 100$ GeV, the quark and antiquark jet rates are very
similar since $x$ is still relatively low.  At higher $p_T$, the valence
contribution increases so that when $p_T = 400$ GeV, the quark rate is 1.5
times the antiquark rate.  Since the initial kinematics are the same for
$\gamma +$jet and jet+jet final states, the average momentum fractions
for $\gamma +$jet production are similar to those shown for
the $\gamma q \rightarrow gq$ channel in 
Fig.~\ref{avexjet}.

The shadowing ratios shown in Fig.~\ref{compdir}(b) are dominated by valence
quarks for $p_T > 100$ GeV.  The FGS ratio is slightly higher because the EKS98
parameterization includes sea quark shadowing.  The effect is similar to the
produced gluon ratios, at the same values of $x$ in Fig.~\ref{jetdir}(c) and
(d), since the final-state gluons can only come from quark and antiquark
induced processes.  

\begin{figure}[htbp]
\setlength{\epsfxsize=0.95\textwidth}
\setlength{\epsfysize=0.55\textheight}
\centerline{\epsffile{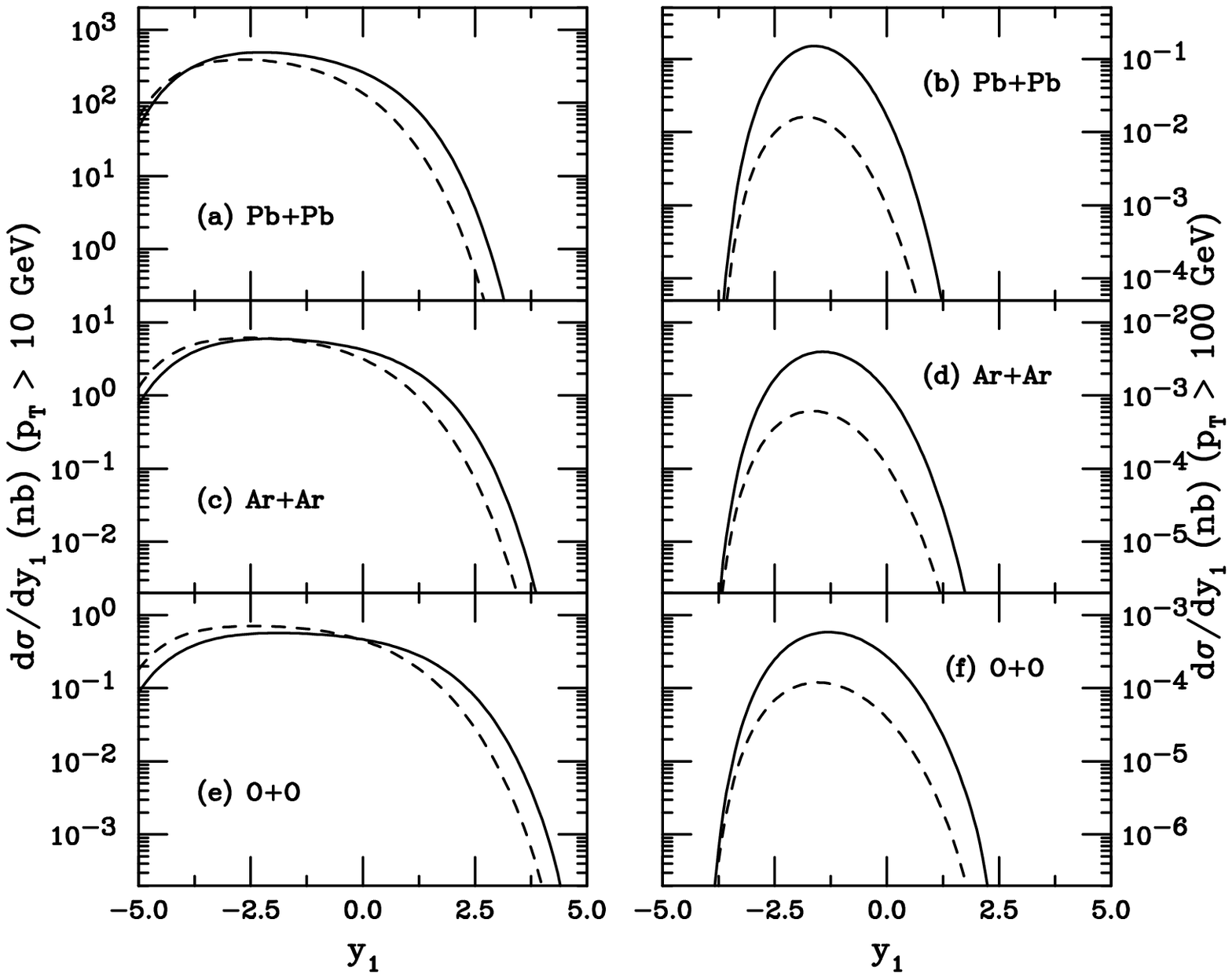}}
\caption[] {We compare the rapidity distributions of direct and resolved
$\gamma+$jet 
photoproduction in peripheral collisions.  The left-hand side shows the results
for $p_T > 10$ GeV for (a) Pb+Pb, (c) Ar+Ar and (e) O+O collisions
while the right-hand side is for $p_T > 100$ GeV for (b) Pb+Pb, (d) Ar+Ar
and (f) O+O collisions.  The solid curves are 
the direct results while the dashed curves 
show the resolved results.  The photon comes from the left.  Note the different
scales on the $y$-axes.}
\label{compt_rap}
\end{figure}

We next present the rapidity distributions for the same two $p_T$ cuts used for
dijet photoproduction in Fig.~\ref{compt_rap}.  Note that the rapidity 
distribution for $p_T > 10$ GeV is broader at negative $y_1$ than the dijet
distributions in Fig.~\ref{jet_dir_rap} because direct dijet
production is dominated by $\gamma g \rightarrow q \overline q$ at these
$p_T$ while the valence distribution entering the
$\gamma q \rightarrow \gamma q$ does not drop
as rapidly at large $x_2$ as the gluon distribution.  When the turnover at
large negative $y_1$ occurs,
it is steeper than for the dijets.  However, it drops even more steeply at 
forward $y_1$ because the quark distribution is
smaller than the gluon at low $x_2$.  When $p_T > 100$ GeV, the $\gamma
+$jet $y_1$ distribution is narrower than the dijets since the
quark distributions drop faster with increasing $x_2$ at high $p_T$.

\begin{figure}[htbp]
\setlength{\epsfxsize=0.95\textwidth}
\setlength{\epsfysize=0.55\textheight}
\centerline{\epsffile{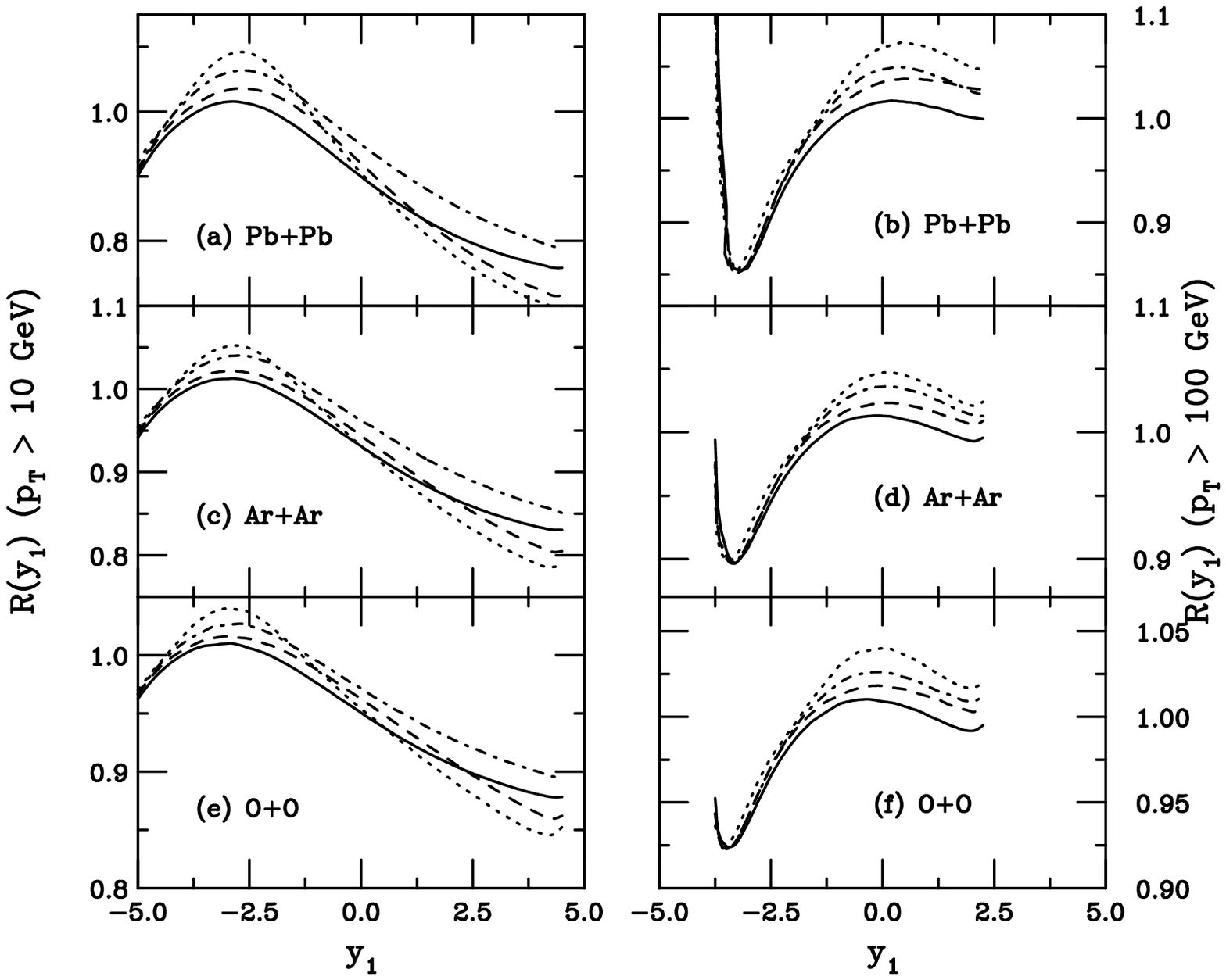}}
\caption[] {We compare shadowing ratios in direct and resolved $\gamma+$jet
production in peripheral collisions. The left-hand side shows the results
for $p_T > 10$ GeV for (a) Pb+Pb, (c) Ar+Ar and (e) O+O collisions
while the right-hand side is for $p_T > 100$ GeV for (b) Pb+Pb, (d) Ar+Ar
and (f) O+O collisions.  The solid and dashed curves give 
the direct ratios for the EKS98 and FGS parameterizations
respectively.  The dot-dashed and dotted curves 
show the resolved ratios for the EKS98 and FGS
parameterizations respectively.}
\label{compt_shad}
\end{figure}

The shadowing ratios as a function of $y_1$ are shown in Fig.~\ref{compt_shad}.
They exhibit some interesting differences from their dijet counterparts in
Fig.~\ref{jet_dir_shad} because of the different production channels.  At
$p_T > 10$ GeV, the antishadowing peak is lower at $y_1 \sim -2.5$ and the
shadowing is larger at $y_1 > 0$.  Although this may seem counterintuitive, 
comparing the valence and sea quark shadowing ratios in Fig.~\ref{shadcomp}
can explain this effect.  Valence antishadowing, the same for EKS98 and FGS,
is not as strong as 
that of the gluon.  The sea quarks have either no antishadowing
(EKS98) or a smaller effect than the valence ratios (FGS).  Thus antishadowing
is reduced for direct $\gamma+$jet production.  At large $y_1$, the $x_2$ 
values, while smaller than those shown in Fig.~\ref{jet_dir_shad}(a) for $|y_1|
\leq 1$, are still moderate.  Since the evolution of the gluon distribution is
faster with $\mu^2$, sea quark shadowing is actually stronger than gluon
shadowing at $p_T > 10$ GeV and low $x_2$.  When $p_T > 100$ GeV, the Fermi
momentum peak is not as prominent because the sharp increase in the valence
and sea shadowing ratios appears at higher $x_2$ than for the gluons, muting
the effect, particularly for the lighter systems.

The cross sections and
rates in a one month, $10^6$ s, run, 
shown in Table~\ref{comdirrates}, are lower than
those for hadroproduction.  Direct $\gamma +$jet
photoproduction proceeds through fewer channels than
hadroproduction where the LO channels are $g q \rightarrow \gamma q$ and
$q \overline q \rightarrow g \gamma$, the same diagrams for resolved
$\gamma+$jet photoproduction.  This, along with the lower effective energy and
correspondingly higher $x$, reduces the photoproduction cross sections relative
to hadroproduction.  The lower $A$ scaling for photoproduction also restricts
the high $p_T$ photoproduction rate.

\subsection{Direct $\gamma+$hadron production}

\begin{figure}[htbp]
\setlength{\epsfxsize=0.95\textwidth}
\setlength{\epsfysize=0.45\textheight}
\centerline{\epsffile{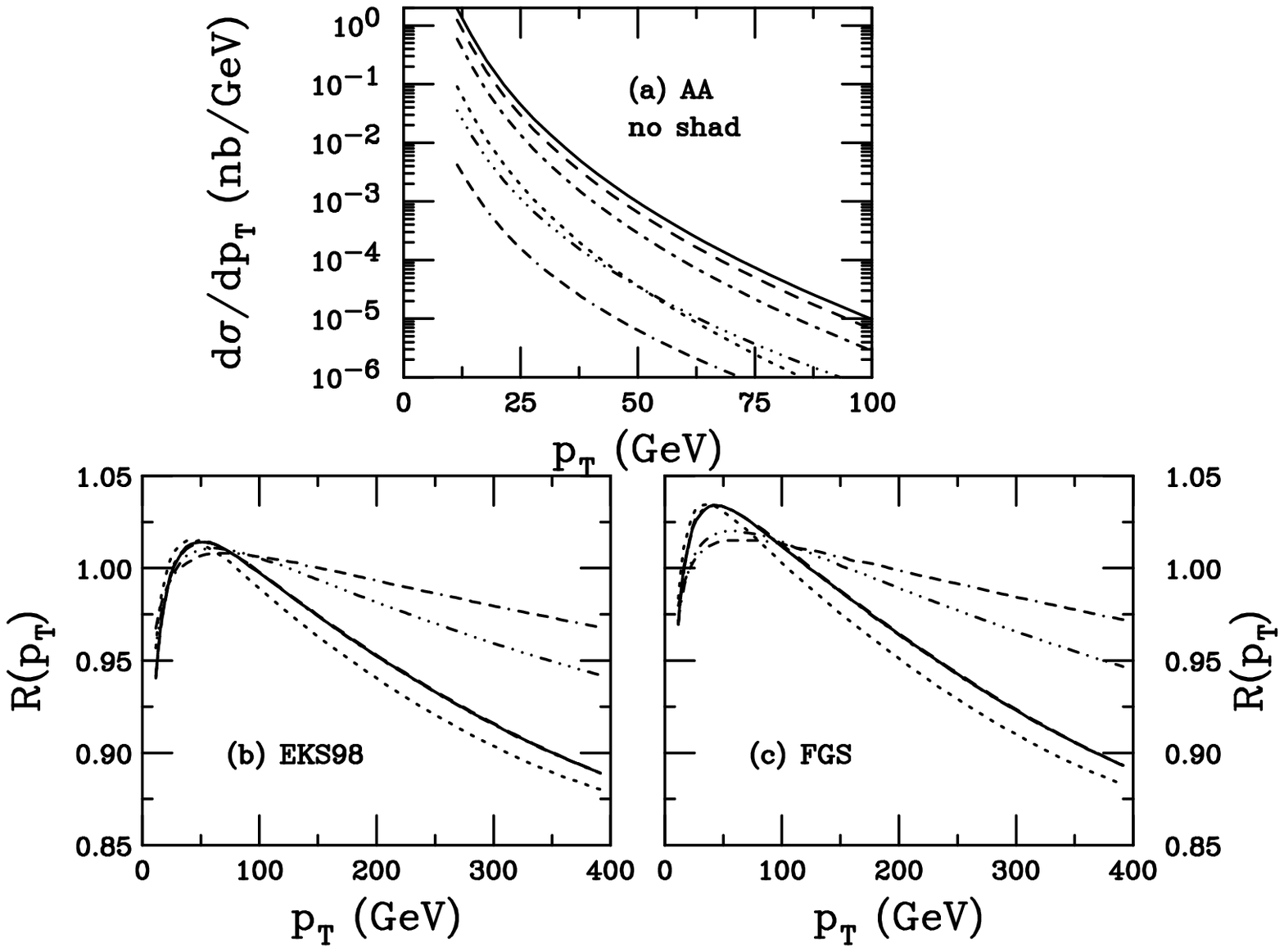}}
\caption[] {Direct leading hadrons from $\gamma+$jet photoproduction 
in peripheral collisions.
(a) The $p_T$ distributions for $|y_1| \leq 1$ are shown for $AA$
collisions.  The Pb+Pb results are shown for charged pions (dashed), kaons 
(dot-dashed), protons (dotted) and the sum of all charged hadrons (solid).   
The charged hadron $p_T$ distributions are also shown for
Ar+Ar (dot-dot-dot-dashed) and O+O (dot-dash-dash-dashed) collisions.
(b) The EKS98 shadowing ratios 
for produced hadrons. The results for pions, kaons and the charged hadron total
(solid) are nearly identical.  The proton result (dotted) is lower.  The total 
charged hadron ratios for Ar+Ar (dot-dot-dot-dashed) and O+O
(dot-dash-dash-dashed) collisions are also shown. (c) The same as (b) for FGS.}
\label{comphaddir}
\end{figure}

We now turn to a description of final-state hadron production opposite a
photon.  The leading particle $p_T$ distribution
is \cite{Field:uq}
\begin{eqnarray}
\label{haddircomp}
\frac{d\sigma_{\gamma A \rightarrow h X}^{\rm dir}}{dp_T d^2b} & = &
4 p_T \int dz \int_{\theta_{\rm min}}^{\theta_{\rm max}} 
\frac{d\theta_{\rm cm}}{\sin
\theta_{\rm cm}} \int dk \frac{d^3N_\gamma}{dk d^2b}
\int \frac{dx_2}{x_2} \\ &  & \mbox{} \times \bigg[ \sum_{i=q, \overline q} 
F_i^A(x_2,\mu^2, \vec b, z) 
\frac{d\sigma_{\gamma i \rightarrow \gamma i}}{dt} 
\frac{D_{h/i}(z_c,\mu^2)}{z_c} \bigg] \, \,   \nonumber
\end{eqnarray}
where $X$ on the left-hand side includes the final-state gluon.
On the partonic level, 
both the initial and final state partons are identical so that parton $i$
fragments into hadron $h$ according to the 
fragmentation function, $D_{h/i}(z_c,\mu^2)$.
The subprocess cross sections, $d\sigma/dt$, 
are related to 
$s^2 d\sigma/dt du$ in Eq.~(\ref{compdireq})
through the momentum-conserving 
delta function $\delta(s + t + u)$ and division by $s^2$.
Our results, shown in Fig.~\ref{comphaddir}, 
are presented in the interval $|y_1| \leq 1$.

The cross section for $\gamma+$hadron 
production are, again, several orders of magnitude
lower than the dijet calculations shown in Fig.~\ref{jethaddir}(a).
At the values of $z_c$ and $x$ important for dijet production, the final state
is dominated by quarks and antiquarks which fragment more frequently into
charged hadrons than do gluons.  While $\gamma g \rightarrow q \overline q$
produces quarks and antiquarks with identical distributions, the contribution 
from the $\gamma q
\rightarrow q g$ channel makes {\it e.g.}\ pion production by quarks and
antiquarks asymmetric.  We also note that for $p_T < 100$ GeV, 60\%
of the dijet
final state particles are pions, $\approx 33$\% kaons and $\approx 7$\%
protons.  As $p_T$ increases, the pion and proton contributions decrease
slightly while the kaon fraction increases.  In the case of $\gamma +$hadron 
final
states, there is no initial state gluon channel.  Thus the valence quarks 
dominate hadron production and the relative fraction of produced pions 
increases to 66\%.  The kaon and proton fractions are subsequently
decreased to $\approx 28$\% and $\approx 6$\% respectively.

The shadowing ratios, shown in Fig.~\ref{comphaddir}(b) and (c) for produced
pions, kaons and protons separately for Pb+Pb as well as the total ratios
for Ar+Ar and O+O collisions, reflect the quark-initiated processes.  We show
the results for all charged hadrons here since we do not differentiate between
quark and antiquark production. The ratios, almost identical for 
produced pions, kaons and charged
hadrons, are quite different from the ratios shown for pion production
by quarks and antiquarks in
Fig.~\ref{jethaddir}(c) and (d) since these pions originate 
from initial-state gluons and thus exhibit
antishadowing. The results are similar to pions from
gluon jets in Fig.~\ref{jethaddir}.  However, in this case 
the
ratios are slightly higher due to the relative couplings.  The proton ratios
are lower than those for pions and kaons due to the nuclear
isospin.  The dominance of $d$ valence quarks in nuclei reduces the proton
production rate since $d$ quarks are only half as effective at producing
protons as $u$ quarks in the KKP fragmentation scheme \cite{Kniehl:2000fe}.
This lower weight in the final state reduces the effectiveness of proton
production by the initial state, decreasing the produced proton shadowing 
ratios relative to pions and kaons.  
Valence quarks dominate the observed final state 
shadowing at these larger values of $x$, as in Fig.~\ref{avexjet}.

\subsection{Resolved $\gamma+$jet production}

Now we turn to resolved production of $\gamma +$jet final states.  
The resolved jet photoproduction cross
section for partons of flavor $f$ in the
subprocess $ij\rightarrow k\gamma$ in $AB$ collisions is modified from 
Eq.~(\ref{sigjetres}) so that now
\begin{eqnarray} S_{NN}^2
\frac{d\sigma^{\rm res}_{\gamma A \rightarrow \gamma \, + \, {\rm 
jet}X}}{dT dU d^2b} & = & 2 \int dz \int_{k_{\rm min}}^\infty 
\frac{dk}{k} {d^3N_\gamma\over dkd^2b} \int_{k_{\rm min}/k}^1 \frac{dx}{x}
\int_{x_{2_{\rm min}}}^1 \frac{dx_2}{x_2} \nonumber \\
&  & \mbox{} \times \sum_{{ij=}\atop{\langle kl \rangle}} \left\{
F_i^\gamma (x,\mu^2) F_j^A(x_2,\mu^2,\vec b,z) + F_j^\gamma (x,\mu^2)
F_i^A(x_2,\mu^2,\vec b,z) \right\} \nonumber \\ 
&  & \mbox{} \times \delta_{fk}
\left[\hat{s}^2\frac{d\sigma^{ij\rightarrow k\gamma}}{d\hat t d\hat u}(\hat t,
\hat u) 
+ \hat{s}^2 \frac{d\sigma^{ij\rightarrow k\gamma}}{d\hat t d\hat u}(\hat u, 
\hat t)
\right] \, \, .
\label{sigcompres} 
\end{eqnarray}
The resolved diagrams are those for hadroproduction of direct photons, $q g
\rightarrow q \gamma$ and $q \overline q \rightarrow g\gamma$.
The $2 \rightarrow 2$ minijet subprocess cross sections are 
\cite{Owens:1986mp}
\begin{eqnarray}
\label{qgtogamq}
\hat s^2 \frac{d^2\sigma_{qg}}{d\hat t d\hat u} & = & - \frac{1}{3} 
\pi \alpha_s \alpha e_Q^2 \bigg[ \frac{\hat s^2 + \hat u^2}{\hat s \hat u}
\bigg] \delta(
\hat s + \hat t + \hat u) \, \, \\
\label{qqbtogamg}
\hat s^2 \frac{d^2\sigma_{q \overline q}}{d\hat t d\hat u} & = & 
\frac{8}{9} \pi \alpha_s \alpha e_Q^2 \bigg[ \frac{\hat t^2 + \hat u^2}{\hat t
\hat u} \bigg] \delta(\hat s + \hat t + \hat u) \, \, .
\end{eqnarray}
Note that
there is no factor $1/(1 + \delta_{kl})$, as in Eq.~(\ref{sigjetres}), since
there are no identical particles in
the final state.

\begin{figure}[htbp]
\setlength{\epsfxsize=0.95\textwidth}
\setlength{\epsfysize=0.45\textheight}
\centerline{\epsffile{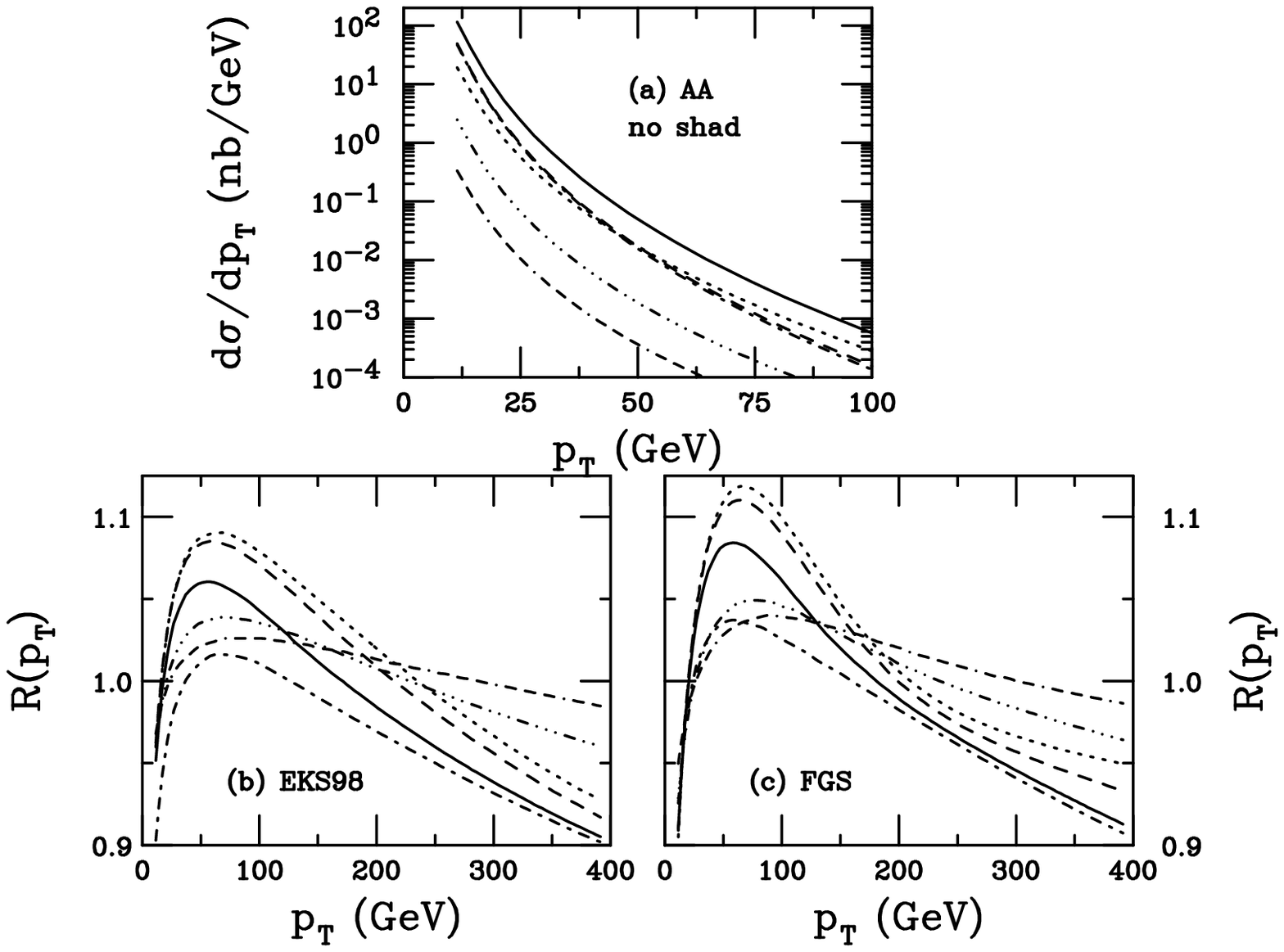}}
\caption[] {Resolved $\gamma +$jet photoproduction in peripheral $AA$ 
collisions. (a)
The Pb+Pb jet $p_T$ distributions with $|y_1| \leq 1$ are shown for quarks 
(dashed), antiquarks (dot-dashed), gluons (dotted) and the total (solid).  
We also show
the total jet $p_T$ distributions in Ar+Ar (dot-dot-dot-dashed) and 
O+O (dash-dash-dash-dotted) collisions. (b) The relative EKS98 shadowing 
contributions 
from quarks (dashed), antiquarks (dotted) and gluons (dot-dashed) as well
as the total (solid) are shown for Pb+Pb collisions.  The totals are also shown
for Ar+Ar (dot-dot-dot-dashed) and O+O (dash-dash-dash-dotted) interactions.}
\label{compjetres}
\end{figure}

The resolved $\gamma+$jet results are shown in Fig.~\ref{compjetres} 
using the GRV LO photon parton densities.  Along with the
total partonic rates in Pb+Pb collisions, we also show the
individual partonic contributions to the jet $p_T$ distributions in
Fig.~\ref{compjetres}(a).  The total yields 
are slightly higher for the resolved
than the direct contribution where only one channel is
open and the coupling is smaller.  
Quark and antiquark production by the $qg$ process
is dominant for $p_T < 40$ GeV but, at higher $p_T$, gluon production dominates
from the $q \overline q$ channel.  The large values of $x$ again
makes the valence
quark contribution dominant at higher $p_T$.
The total $p_T$ distributions
for Ar+Ar and O+O collisions are also shown.

The strong antishadowing in the produced quark and antiquark ratios 
in Fig.~\ref{compjetres}(b) and (c) comes from the $qg$
channel.  The antiquark ratio is higher because the $qg$ parton luminosity
peaks at higher $x$ than the $\overline q g$ luminosity and at lower $x$ the
gluon antishadowing ratio is larger.  The difference between the
quark and antiquark ratios increases with $p_T$ since the average $x$ 
and thus the valence quark
contribution also grow with $p_T$.  At high $p_T$, the
flattening of the FGS quark and antiquark ratios is due to the
flattening of the gluon parameterization at $x > 0.2$.  

The final-state gluon ratio shows little antishadowing since it
arises from the $q \overline q$ channel.  The antishadowing in the EKS98 ratio
is due to the valence quarks while the higher ratio for FGS reflects the fact
that the antiquark ratios also show antishadowing for $x < 0.2$.  The ratio
for the total is essentially the average of the three 
contributions at low $p_T$, where they are similar, while at high $p_T$, 
where the $q \overline q$
channel dominates, the total ratio approximates the produced gluon
ratio in both cases.

The resolved rapidity distributions are also shown in Fig.~\ref{compt_rap} for
the two $p_T$ cuts.  The resolved distribution is not as broad at negative
$y_1$ as that of the dijet process in Fig.~\ref{jet_dir_rap} due to the 
smaller relative gluon contribution and the reduced number of channels 
available for the $\gamma+$jet process.  Note that the relative resolved to 
direct production is reduced here and the direct process is actually dominant
at $y_1 >0$ for $p_T > 10$ GeV and for all $y_1$ at $p_T > 100$ GeV.
The antishadowing peak is higher for resolved production, shown in 
Fig.~\ref{compt_shad}, thanks to the gluon contribution to resolved production.

\begin{figure}[htbp]
\setlength{\epsfxsize=0.95\textwidth}
\setlength{\epsfysize=0.3\textheight}
\centerline{\epsffile{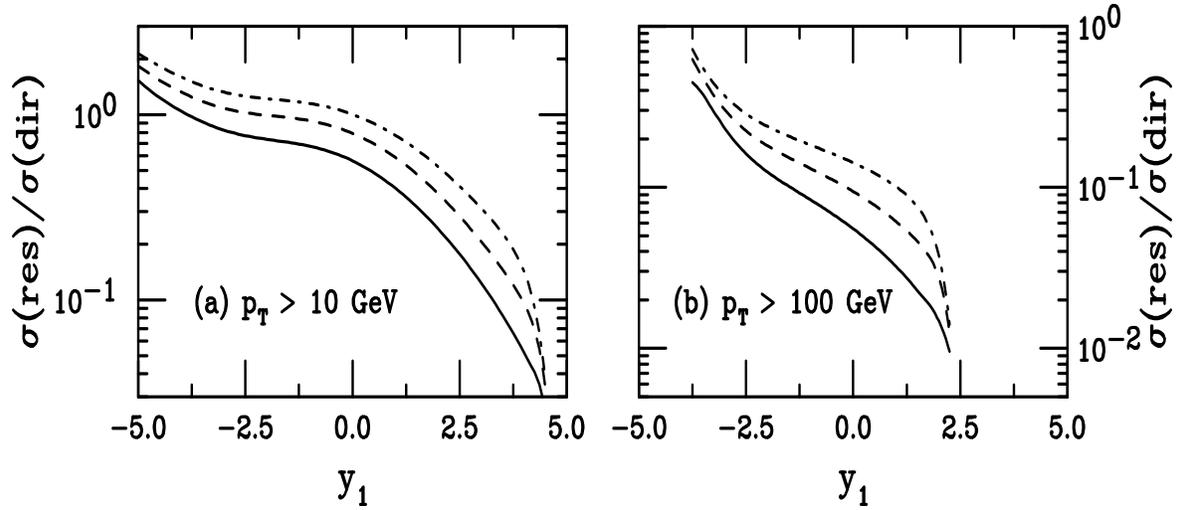}}
\caption[] {We present the resolved/direct $\gamma+$jet production ratios as a 
function of rapidity. The left-hand side shows the results
for $p_T > 10$ GeV
while the right-hand side is for $p_T > 100$ GeV.  
The curves are Pb+Pb (solid), Ar+Ar (dashed) and O+O (dot-dashed).  The photon
comes from the left.}
\label{compt_rat}
\end{figure}

Finally, we show the resolved to direct ratio in Fig.~\ref{compt_rat}.  The
direct rate alone should be observable at $y_1 > -4$ for Pb+Pb, $y_1 \sim -2.5$
for Ar+Ar and $y_1 \sim 0$ 
for O+O and $p_T > 10$ GeV.  Direct production dominates over
all $y_1$ by a large factor when $p_T > 100$ GeV.  Although the rates are 
lower than the dijet rates, the dominance of direct $\gamma+$jet production
implies than the nuclear quark distribution can be cleanly studied.

The resolved $\gamma +$jet rates are shown in Table~\ref{comresrates}.  These
rates are only slightly larger than 
the direct rates for $p_T \leq 20$ GeV but drop
below the direct rates at higher $p_T$ since there is no large growth 
in the number of available production channels, as in dijet production.
In addition, the lower effective energy plays an important role here as well. 

\subsection{Resolved $\gamma+$hadron production}

The leading particle $p_T$ distributions of jets from $\gamma+$jet production
are
\begin{eqnarray}
\frac{d\sigma^{\rm res}_{\gamma A \rightarrow\gamma+ hX}}{dp_T d^2b} & = & 
4p_T \int dz  
\int_{\theta_{\rm min}}^{\theta_{\rm max}}
\frac{d\theta_{\rm cm}}{\sin \theta_{\rm cm}}
\int_{k_{\rm min}}^\infty 
\frac{dk}{k} {d^3N_\gamma\over dkd^2b} \int_{k_{\rm min}/k}^1 \frac{dx}{x}
\int_{x_{2_{\rm min}}}^1 \frac{dx_2}{x_2} \nonumber \\
&  & \mbox{} \times \sum_{{ij=}\atop{\langle kl \rangle}} \left\{
F_i^\gamma (x,\mu^2) F_j^A(x_2,\mu^2,\vec b,z) + F_j^\gamma (x,\mu^2)
F_i^A(x_2,\mu^2,\vec b,z) \right\} \nonumber \\ 
&  & \mbox{} \times \delta_{fk}
\left[ {d\sigma\over d\hat t}^{ij\rightarrow k\gamma}(\hat t, \hat
u) 
+ {d\sigma\over d\hat t}^{ij\rightarrow k\gamma}
(\hat u, \hat t)
\right] \frac{D_{h/k}(z_c,\mu^2)}{z_c} \, \, .
\label{sigcomphadres} 
\end{eqnarray}
The subprocess cross sections, $d\sigma/d\hat t$, 
are related to 
$\hat s^2 d\sigma/d\hat t d\hat u$ in Eq.~(\ref{sigcompres})
through the momentum-conserving 
delta function $\delta(\hat s + \hat t + \hat u)$ and division by $\hat s^2$.

\begin{figure}[htbp]
\setlength{\epsfxsize=0.95\textwidth}
\setlength{\epsfysize=0.45\textheight}
\centerline{\epsffile{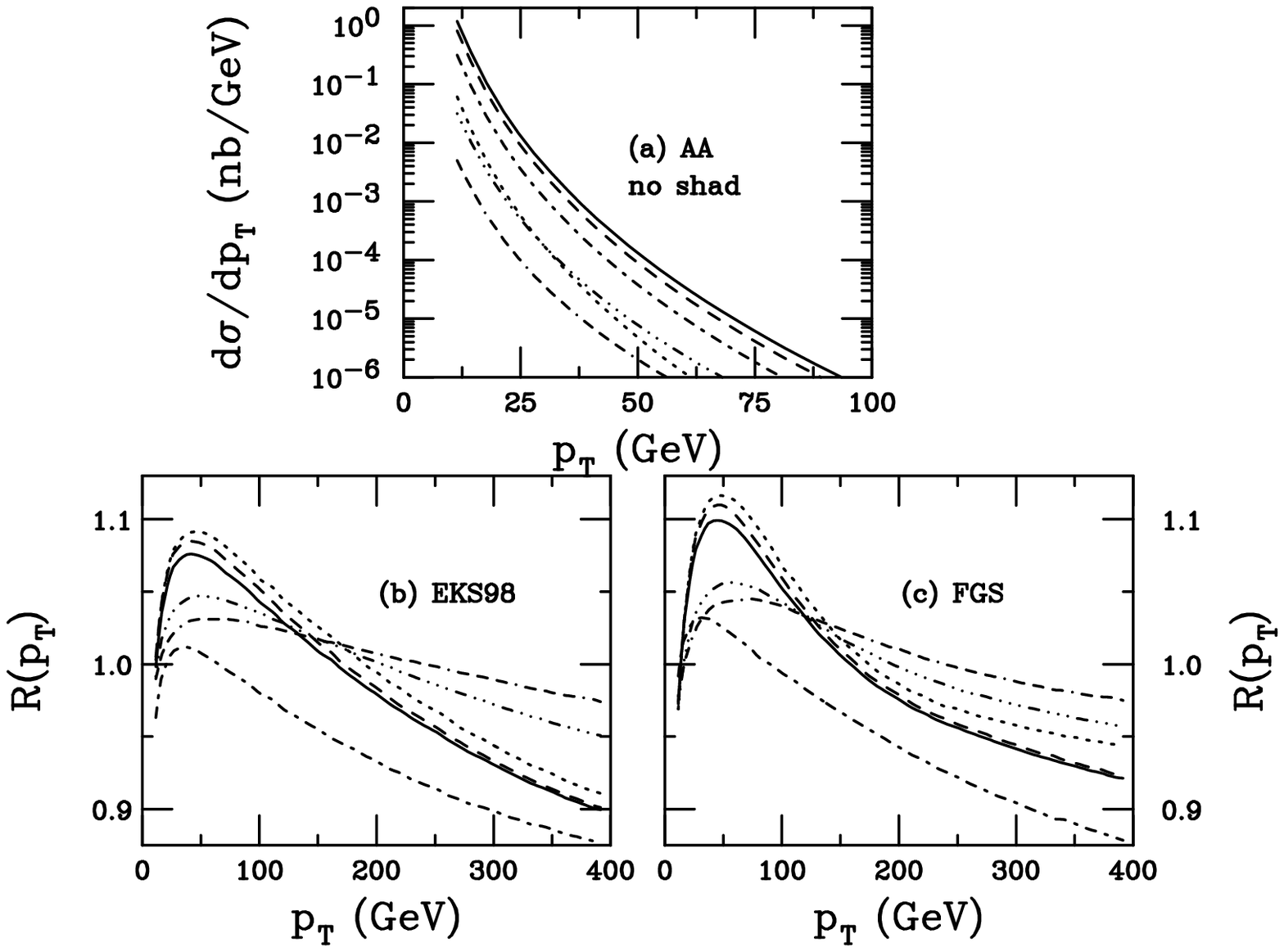}}
\caption[] {Resolved leading hadrons from $\gamma +$jet
photoproduction in peripheral collisions.
(a) The $p_T$ distributions for $|y_1| \leq 1$ are shown for $AA$
collisions.  The Pb+Pb results are shown for charged pions (dashed), kaons 
(dot-dashed), protons (dotted) and the sum of all charged hadrons (solid).   
The charged hadron $p_T$ distributions are also shown for
Ar+Ar (dot-dot-dot-dashed) and O+O (dot-dash-dash-dashed) collisions.
(b) The EKS98 shadowing ratios for produced pions.  For Pb+Pb collisions,
we show the ratios for pions produced by quarks (dashed), antiquarks 
(dotted), gluons (dot-dashed) and the total (solid) separately.  The ratios 
for pions produced by all partons are also shown for Ar+Ar (dot-dot-dot-dashed)
and O+O (dot-dash-dash-dashed) collisions. (c) The same as (b) for FGS.
}
\label{comphadres}
\end{figure}

The resolved $p_T$ distributions for hadrons
are shown in Fig.~\ref{comphadres}(a).  
Note that the resolved cross section for leading hadron production is
similar to direct production, shown in Fig.~\ref{comphaddir}(a).
The same effect is seen for dijet production in Figs.~\ref{jethadres} and
\ref{jethaddir}.  

The shadowing ratios are shown in Fig.~\ref{comphadres}.  
The difference between the shadowing ratios
for pions produced by quarks and antiquarks is rather large and reflects both
gluon antishadowing at low $p_T$ as well as the relative valence to sea
contributions for quark and antiquark production through $q(\overline q) g
\rightarrow q(\overline q) \gamma$.  In the FGS calculations, the
antiquark ratio reflects the flattening of the antiquark and gluon ratios
at $x > 0.2$.  Since pions produced by gluons come from the $q \overline q
\rightarrow \gamma g$ channel alone, only a small effect is seen, primarily in
the EMC region.  Now the total pion rates follow those for quark and antiquark
production of final-state pions instead of those of the gluon.

Although our $p_T$-dependent calculations have focused on the midrapidity
region of $|y_1|\leq 1$, we have shown that extending the rapidity coverage 
could lead to greater sensitivity to the small $x_2$ region and larger
contributions from direct photoproduction, especially at low $p_T$.  

Thus $\gamma +$jet production is a good way to measure the nuclear
quark distribution functions.  Direct photoproduction is dominant 
at central rapidities for moderate values of $p_T$.  
Final-state hadron production is somewhat larger for 
direct production so that, even
if the rates are low, the results will be relatively clean.

\section{Summary}

There are a number of uncertainties in our results.  All our
calculations are at leading order so that there is some uncertainty in the
total rate, see Refs.~\cite{Klein:2002wm,Vogt:2002eu}.  
Some uncertainty also arises from the
scale dependence, both in the parton densities and in the fragmentation 
functions.  The fragmentation functions at large $z_c$ also introduce
uncontrollable uncertainties in the rates.  Hopefully more data will bring
the parton densities in the photon, proton and nucleus under better control
before the LHC begins operation.  The data from RHIC also promises to bring the
fragmentation functions under better control in the near future.

While the photon flux is also an uncertainty, it can be determined
experimentally.  The hadronic interaction
probability near the minimum radius depends on the matter distribution
in the nucleus.  Our calculations use Woods-Saxon distributions with
parameters fit to electron scattering data.  This data is quite
accurate.  However, electron scattering is only sensitive to the
charge distribution in the nucleus.  Recent measurements indicate that
the neutron and proton distributions differ in nuclei \cite{Trzcinska:sy}.
This uncertainty in the matter distribution is likely to limit the
photon flux determination.

The uncertainty in the photon flux can be reduced by
calibrating it with other measurements such as vector meson production,
$\gamma A \rightarrow VA$.  
Studies of well known two-photon processes, like lepton production,
can also help refine the determination of the photon flux.  With such
checks, it should be possible to understand the photon flux in $pA$
relative to $AA$ to better than 10\%, good enough for a useful shadowing
measurement.

We would like to thank S. R. Klein and M. Strikman for discussions.
This work was
supported in part by the Division of Nuclear Physics of the Office of
High Energy and Nuclear Physics of the U.S. Department of Energy under
Contract No. DE-AC-03-76SF00098.

\begin{table}
\begin{center}
\begin{tabular}{cccccccc}
$A$       & $L$ (mb$^{-1}$s$^{-1}$) & $\sqrt{S_{NN}}$ (GeV) & $E_{\rm beam}$ 
(GeV) & $\gamma_L$ & $k_{\rm max}$ (GeV) & $E_{\rm max}$ (TeV) 
& $\sqrt{S_{\gamma N}}$ (GeV) 
\\ \hline
O    & 160  & 7000 & 3500 & 3730 & 243 & 1820 & 1850 \\
Ar   & 43   & 6300 & 3150 & 3360 & 161 & 1080 & 1430 \\
Pb   & 0.42 & 5500 & 2750 & 2930 &  81 &  480 &  950 \\ \hline
\end{tabular}
\end{center}
\caption[]{Pertinent parameters and kinematic limits for $AA$
collisions at the LHC {\protect \cite{Klein:2002wm}}.  
We first give the luminosities and the $NN$ collision
kinematics,  the nucleon-nucleon center of mass energies, $\sqrt{S_{NN}}$, the
corresponding beam energies, $E_{\rm beam}$, Lorentz factors, $\gamma_L$.  
We then present the
photon cutoff energies in the center of mass frame, $k_{\rm max}$, and
in the nuclear rest frame, $E_{\rm max}$, as well as the equivalent maximum
photon-nucleon center of mass energies, $\sqrt{S_{\gamma N}}$.}
\label{gamfacs}
\end{table}

\begin{table}
\begin{center}
\begin{tabular}{c|cc||cc||cc}
& \multicolumn{2}{c||}{Pb+Pb} & \multicolumn{2}{c||}{Ar+Ar} &
\multicolumn{2}{c}{O+O} \\
$p_T$ (GeV) & $d\sigma/dp_T$ & Rate & $d\sigma/dp_T$ & Rate &
$d\sigma/dp_T$ & Rate  \\ \hline
 11.4 & 2.802$\times 10^5$ & 1.177$\times 10^5$ & 4.771$\times 10^3$ & 
2.052$\times 10^5$ & 5.506$\times 10^2$ & 8.810$\times 10^4$ \\
 21.4 & 1.389$\times 10^4$ & 5.836$\times 10^3$ & 2.869$\times 10^2$ &
 1.234$\times 10^4$ & 3.706$\times 10^1$ & 5.930$\times 10^3$ \\
 31.4 & 1.907$\times 10^3$ & 8.010$\times 10^2$ & 4.574$\times 10^1$ &
 1.967$\times 10^3$ & 6.462$\times 10^0$ & 1.034$\times 10^3$ \\
 41.4 & 4.181$\times 10^2$ & 1.756$\times 10^2$ & 1.137$\times 10^1$ &
 4.889$\times 10^2$ & 1.734$\times 10^0$ & 2.774$\times 10^2$ \\
 51.4 & 1.207$\times 10^2$ & 5.069$\times 10^1$ & 3.668$\times 10^0$ &
 1.577$\times 10^2$ & 5.983$\times 10^{-1}$ & 9.573$\times 10^1$ \\
 61.4 & 4.172$\times 10^1$ & 1.752$\times 10^1$ & 1.403$\times 10^0$ &
 6.033$\times 10^1$ & 2.434$\times 10^{-1}$ & 3.894$\times 10^1$ \\
 71.4 & 1.640$\times 10^1$ & 6.890$\times 10^0$ & 6.061$\times 10^{-1}$ & 
 2.606$\times 10^1$ & 1.113$\times 10^{-1}$ & 1.781$\times 10^1$ \\
 81.4 & 7.105$\times 10^0$ & 2.984$\times 10^0$ & 2.868$\times 10^{-1}$ & 
 1.233$\times 10^1$ & 5.557$\times 10^{-2}$ & 8.891$\times 10^0$ \\
 91.4 & 3.316$\times 10^0$ & 1.393$\times 10^0$ & 1.457$\times 10^{-1}$ & 
 6.265$\times 10^0$ & 2.970$\times 10^{-2}$ & 4.752$\times 10^0$ \\
 101  & 1.640$\times 10^0$ & 6.888$\times 10^{-1}$ & 7.832$\times 10^{-2}$ & 
 3.368$\times 10^0$ & 1.676$\times 10^{-2}$ & 2.682$\times 10^0$ \\
 111  & 8.538$\times 10^{-1}$ & 3.586$\times 10^{-1}$ & 4.416$\times 10^{-2}$ 
 & 1.899$\times 10^0$ & 9.898$\times 10^{-3}$ & 1.584$\times 10^0$ \\
 121  & 4.624$\times 10^{-1}$ & 1.942$\times 10^{-1}$ & 2.586$\times 10^{-2}$ &
 1.112$\times 10^0$ & 6.066$\times 10^{-3}$ & 9.706$\times 10^{-1}$  \\
 131  & 2.588$\times 10^{-1}$ & 1.087$\times 10^{-1}$ & 1.564$\times 10^{-2}$ &
 6.725$\times 10^{-1}$ & 3.835$\times 10^{-3}$ & 6.136$\times 10^{-1}$ \\
 141  & 1.493$\times 10^{-1}$ & 6.270$\times 10^{-2}$ & 9.737$\times 10^{-3}$ &
 4.187$\times 10^{-1}$ & 2.493$\times 10^{-3}$ & 3.989$\times 10^{-1}$ \\
 151  & 8.838$\times 10^{-2}$ & 3.712$\times 10^{-2}$ & 6.215$\times 10^{-3}$ &
 2.672$\times 10^{-1}$ & 1.660$\times 10^{-3}$ & 2.656$\times 10^{-1}$ \\
 161  & 5.346$\times 10^{-2}$ & 2.245$\times 10^{-2}$ & 4.050$\times 10^{-3}$ &
 1.741$\times 10^{-1}$ & 1.128$\times 10^{-3}$ & 1.805$\times 10^{-1}$ \\
 171  & 3.293$\times 10^{-2}$ & 1.383$\times 10^{-2}$ & 2.687$\times 10^{-3}$ &
 1.155$\times 10^{-1}$ & 7.803$\times 10^{-4}$ & 1.248$\times 10^{-1}$ \\
 181  & 2.064$\times 10^{-2}$ & 8.669$\times 10^{-3}$ & 1.813$\times 10^{-3}$ &
 7.796$\times 10^{-2}$ & 5.485$\times 10^{-4}$ & 8.776$\times 10^{-2}$ \\
 191  & 1.316$\times 10^{-2}$ & 5.528$\times 10^{-3}$ & 1.244$\times 10^{-3}$ &
 5.349$\times 10^{-2}$ & 3.917$\times 10^{-4}$ & 6.267$\times 10^{-2}$ \\
 201  & 8.507$\times 10^{-3}$ & 3.573$\times 10^{-3}$ & 8.642$\times 10^{-4}$ &
 3.716$\times 10^{-2}$ & 2.833$\times 10^{-4}$ & 4.533$\times 10^{-2}$ \\
 \hline 
\end{tabular}
\caption[]{The differential cross sections, in units of nb/GeV, 
without shadowing and rates for $|y_1| \leq 1$
in a one month, $10^6$ s, run
for direct dijet photoproduction in peripheral $AA$ collisions.}
\label{jetdirrates}
\end{center}
\end{table}

\begin{table}
\begin{center}
\begin{tabular}{c|cc||cc||cc}
& \multicolumn{2}{c||}{Pb+Pb} & \multicolumn{2}{c||}{Ar+Ar} &
\multicolumn{2}{c}{O+O} \\
$p_T$ (GeV) & $d\sigma/dp_T$ & Rate & $d\sigma/dp_T$ & Rate &
$d\sigma/dp_T$ & Rate  \\ \hline
 11.4 & 5.184$\times 10^5$ & 2.177$\times 10^5$ & 1.185$\times 10^4$ & 
5.096$\times 10^5$ & 1.747$\times 10^3$ & 2.795$\times 10^5$ \\
 21.4 & 1.733$\times 10^4$ & 7.279$\times 10^3$ & 4.713$\times 10^2$ & 
2.027$\times 10^4$ & 7.715$\times 10^1$ & 1.234$\times 10^4$ \\
 31.4 & 1.895$\times 10^3$ & 7.959$\times 10^2$ & 5.927$\times 10^1$ & 
2.549$\times 10^3$ & 1.054$\times 10^1$ & 1.686$\times 10^3$ \\
 41.4 & 3.530$\times 10^2$ & 1.483$\times 10^2$ & 1.247$\times 10^1$ & 
5.362$\times 10^2$ & 2.386$\times 10^0$ & 3.818$\times 10^2$ \\
 51.4 & 8.945$\times 10^1$ & 3.757$\times 10^1$ & 3.530$\times 10^0$ & 
1.518$\times 10^2$ & 7.210$\times 10^{-1}$ & 1.154$\times 10^2$ \\
 61.4 & 2.767$\times 10^1$ & 1.162$\times 10^1$ & 1.210$\times 10^0$ & 
5.203$\times 10^1$ & 2.629$\times 10^{-1}$ & 4.206$\times 10^1$ \\
 71.4 & 9.857$\times 10^0$ & 4.140$\times 10^0$ & 4.754$\times 10^{-1}$ & 
2.044$\times 10^1$ & 1.094$\times 10^{-1}$ & 1.750$\times 10^1$ \\
 81.4 & 3.905$\times 10^0$ & 1.640$\times 10^0$ & 2.068$\times 10^{-1}$ & 
8.892$\times 10^0$ & 5.031$\times 10^{-2}$ & 8.050$\times 10^0$ \\
 91.4 & 1.679$\times 10^0$ & 7.052$\times 10^{-1}$ & 9.726$\times 10^{-2}$ & 
4.182$\times 10^0$ & 2.496$\times 10^{-2}$ & 3.994$\times 10^0$ \\
101   & 7.687$\times 10^{-1}$ & 3.229$\times 10^{-1}$ & 4.864$\times 10^{-2}$ &
2.092$\times 10^0$ & 1.315$\times 10^{-2}$ & 2.104$\times 10^0$ \\
111   & 3.725$\times 10^{-1}$ & 1.565$\times 10^{-1}$ & 2.566$\times 10^{-2}$ &
1.103$\times 10^0$ & 7.296$\times 10^{-3}$ & 1.167$\times 10^0$ \\
121   & 1.884$\times 10^{-1}$ & 7.913$\times 10^{-2}$ & 1.411$\times 10^{-2}$ &
6.067$\times 10^{-1}$ & 4.215$\times 10^{-3}$ & 6.744$\times 10^{-1}$ \\
131   & 9.870$\times 10^{-2}$ & 4.145$\times 10^{-2}$ & 8.036$\times 10^{-3}$ &
3.455$\times 10^{-1}$ & 2.519$\times 10^{-3}$ & 4.030$\times 10^{-1}$ \\
141   & 5.353$\times 10^{-2}$ & 2.248$\times 10^{-2}$ & 4.728$\times 10^{-3}$ &
2.033$\times 10^{-1}$ & 1.554$\times 10^{-3}$ & 2.486$\times 10^{-1}$ \\
151   & 2.986$\times 10^{-2}$ & 1.254$\times 10^{-2}$ & 2.858$\times 10^{-3}$ &
1.229$\times 10^{-1}$ & 9.840$\times 10^{-4}$ & 1.574$\times 10^{-1}$ \\
161   & 1.704$\times 10^{-2}$ & 7.157$\times 10^{-3}$ & 1.767$\times 10^{-3}$ &
7.598$\times 10^{-2}$ & 6.372$\times 10^{-4}$ & 1.020$\times 10^{-1}$ \\
171   & 9.929$\times 10^{-3}$ & 4.170$\times 10^{-3}$ & 1.115$\times 10^{-3}$ &
4.795$\times 10^{-2}$ & 4.207$\times 10^{-4}$ & 6.731$\times 10^{-2}$ \\
181   & 5.895$\times 10^{-3}$ & 2.476$\times 10^{-3}$ & 7.159$\times 10^{-4}$ &
3.078$\times 10^{-2}$ & 2.827$\times 10^{-4}$ & 4.523$\times 10^{-2}$ \\
191   & 3.569$\times 10^{-3}$ & 1.499$\times 10^{-3}$ & 4.685$\times 10^{-4}$ &
2.015$\times 10^{-2}$ & 1.935$\times 10^{-4}$ & 3.096$\times 10^{-2}$ \\
201   & 2.192$\times 10^{-3}$ & 9.206$\times 10^{-4}$ & 3.110$\times 10^{-4}$ &
1.337$\times 10^{-2}$ & 1.342$\times 10^{-4}$ & 2.147$\times 10^{-2}$ \\
\hline
\end{tabular}
\caption[]{The differential cross sections, in units of nb/GeV, 
without shadowing and rates with $|y_1| \leq 1$
in a one month, $10^6$ s, run
for resolved dijet photoproduction in peripheral $AA$ collisions.}
\label{jetresrates}
\end{center}
\end{table}

\begin{table}
\begin{center}
\begin{tabular}{c|cc||cc||cc}
& \multicolumn{2}{c||}{Pb+Pb} & \multicolumn{2}{c||}{Ar+Ar} &
\multicolumn{2}{c}{O+O} \\
$p_T$ (GeV) & $d\sigma/dp_T$ & Rate & $d\sigma/dp_T$ & Rate &
$d\sigma/dp_T$ & Rate  \\ \hline
 11.4 & 1.877$\times 10^2$ & 7.886$\times 10^1$ & 2.838$\times 10^0$ & 
 1.220$\times 10^1$ & 3.024$\times 10^{-1}$ & 4.838$\times 10^1$ \\
 21.4 & 1.435$\times 10^1$ & 6.026$\times 10^0$ & 2.534$\times 10^{-1}$ & 
 1.090$\times 10^1$ & 2.948$\times 10^{-2}$ & 4.716$\times 10^0$ \\
 31.4 & 1.656$\times 10^0$ & 1.116$\times 10^0$ & 5.310$\times 10^{-2}$ & 
 2.284$\times 10^0$ & 6.630$\times 10^{-3}$ & 1.061$\times 10^0$ \\
 41.4 & 7.326$\times 10^{-1}$ & 3.076$\times 10^{-1}$ & 1.633$\times 10^{-2}$ &
 7.022$\times 10^{-1}$ & 2.168$\times 10^{-3}$ & 3.468$\times 10^{-1}$ \\
 51.4 & 2.546$\times 10^{-1}$ & 1.069$\times 10^{-1}$ & 6.264$\times 10^{-3}$ &
 2.694$\times 10^{-1}$ & 8.804$\times 10^{-4}$ & 1.408$\times 10^{-1}$ \\
 61.4 & 1.027$\times 10^{-1}$ & 4.314$\times 10^{-2}$ & 2.778$\times 10^{-3}$ &
 1.195$\times 10^{-1}$ & 4.116$\times 10^{-4}$ & 6.586$\times 10^{-2}$ \\
 71.4 & 4.602$\times 10^{-2}$ & 1.933$\times 10^{-2}$ & 1.362$\times 10^{-3}$ &
 5.856$\times 10^{-2}$ & 2.124$\times 10^{-4}$ & 3.400$\times 10^{-2}$ \\
 81.4 & 2.228$\times 10^{-2}$ & 9.358$\times 10^{-3}$ & 7.200$\times 10^{-4}$ &
 3.096$\times 10^{-2}$ & 1.179$\times 10^{-4}$ & 1.887$\times 10^{-2}$ \\
 91.4 & 1.146$\times 10^{-2}$ & 4.814$\times 10^{-3}$ & 4.034$\times 10^{-4}$ &
 1.735$\times 10^{-2}$ & 6.930$\times 10^{-5}$ & 1.109$\times 10^{-2}$ \\
101   & 6.176$\times 10^{-3}$ & 2.594$\times 10^{-3}$ & 2.366$\times 10^{-4}$ &
1.017$\times 10^{-2}$ & 4.258$\times 10^{-5}$ & 6.812$\times 10^{-3}$ \\
111   & 3.464$\times 10^{-3}$ & 1.455$\times 10^{-3}$ & 1.441$\times 10^{-4}$ &
6.196$\times 10^{-3}$ & 2.716$\times 10^{-5}$ & 4.346$\times 10^{-3}$ \\
121   & 2.006$\times 10^{-3}$ & 8.426$\times 10^{-4}$ & 9.052$\times 10^{-5}$ &
3.892$\times 10^{-3}$ & 1.785$\times 10^{-5}$ & 2.856$\times 10^{-3}$ \\
131   & 1.192$\times 10^{-3}$ & 5.004$\times 10^{-4}$ & 5.832$\times 10^{-5}$ &
2.508$\times 10^{-3}$ & 1.203$\times 10^{-5}$ & 1.924$\times 10^{-3}$ \\
141   & 7.254$\times 10^{-4}$ & 3.046$\times 10^{-4}$ & 3.846$\times 10^{-5}$ &
1.654$\times 10^{-3}$ & 8.292$\times 10^{-6}$ & 1.327$\times 10^{-3}$ \\
151   & 4.506$\times 10^{-4}$ & 1.893$\times 10^{-4}$ & 2.586$\times 10^{-5}$ &
1.112$\times 10^{-3}$ & 5.826$\times 10^{-6}$ & 9.322$\times 10^{-4}$ \\
161   & 2.848$\times 10^{-4}$ & 1.196$\times 10^{-4}$ & 1.769$\times 10^{-5}$ &
7.604$\times 10^{-4}$ & 4.160$\times 10^{-6}$ & 6.656$\times 10^{-4}$ \\
171   & 1.825$\times 10^{-4}$ & 7.666$\times 10^{-5}$ & 1.227$\times 10^{-5}$ &
5.274$\times 10^{-4}$ & 3.014$\times 10^{-6}$ & 4.822$\times 10^{-4}$ \\
181   & 1.186$\times 10^{-4}$ & 4.982$\times 10^{-5}$ & 8.620$\times 10^{-6}$ &
3.706$\times 10^{-4}$ & 2.210$\times 10^{-6}$ & 3.536$\times 10^{-4}$ \\
191   & 7.816$\times 10^{-5}$ & 3.282$\times 10^{-5}$ & 6.134$\times 10^{-6}$ &
2.638$\times 10^{-4}$ & 1.642$\times 10^{-6}$ & 2.628$\times 10^{-4}$ \\
201   & 5.204$\times 10^{-5}$ & 2.186$\times 10^{-5}$ & 4.412$\times 10^{-6}$ &
1.897$\times 10^{-4}$ & 1.233$\times 10^{-6}$ & 1.972$\times 10^{-4}$ \\
\hline
\end{tabular}
\caption[]{The differential cross sections, in units of nb/GeV, 
without shadowing and rates in the interval
$|y_1| \leq 1$ in a one month, $10^6$ s, run
for direct $\gamma+$jet photoproduction in peripheral $AA$ collisions.}
\label{comdirrates}
\end{center}
\end{table}

\begin{table}
\begin{center}
\begin{tabular}{c|cc||cc||cc}
& \multicolumn{2}{c||}{Pb+Pb} & \multicolumn{2}{c||}{Ar+Ar} &
\multicolumn{2}{c}{O+O} \\
$p_T$ (GeV) & $d\sigma/dp_T$ & Rate & $d\sigma/dp_T$ & Rate &
$d\sigma/dp_T$ & 
Rate  \\ \hline
 11.4 & 1.154$\times 10^2$ & 4.847$\times 10^1$ & 2.456$\times 10^0$ & 
1.056$\times 10^2$ & 3.333$\times 10^{-1}$ & 5.333$\times 10^1$ \\
 21.4 & 5.352$\times 10^0$ & 2.248$\times 10^0$ & 1.391$\times 10^{-1}$ & 
5.981$\times 10^0$ & 2.137$\times 10^{-2}$ & 3.419$\times 10^0$ \\
 31.4 & 7.037$\times 10^{-1}$ & 2.956$\times 10^{-1}$ & 2.126$\times 10^{-2}$ &
 9.142$\times 10^{-1}$ & 3.588$\times 10^{-3}$ & 5.741$\times 10^{-1}$ \\
 41.4 & 1.497$\times 10^{-1}$ & 6.287$\times 10^{-2}$ & 5.123$\times 10^{-3}$ &
 2.203$\times 10^{-1}$ & 9.352$\times 10^{-4}$ & 1.496$\times 10^{-1}$ \\
 51.4 & 4.216$\times 10^{-2}$ & 1.771$\times 10^{-2}$ & 1.611$\times 10^{-3}$ &
 6.927$\times 10^{-2}$ & 3.151$\times 10^{-4}$ & 5.042$\times 10^{-2}$ \\
 61.4 & 1.427$\times 10^{-2}$ & 5.993$\times 10^{-3}$ & 6.037$\times 10^{-4}$ &
 2.596$\times 10^{-2}$ & 1.257$\times 10^{-4}$ & 2.011$\times 10^{-2}$ \\
 71.4 & 5.502$\times 10^{-3}$ & 2.311$\times 10^{-3}$ & 2.561$\times 10^{-4}$ &
 1.101$\times 10^{-2}$ & 5.647$\times 10^{-5}$ & 9.035$\times 10^{-3}$ \\
 81.4 & 2.338$\times 10^{-3}$ & 9.820$\times 10^{-4}$ & 1.192$\times 10^{-4}$ &
 5.126$\times 10^{-3}$ & 2.776$\times 10^{-5}$ & 4.442$\times 10^{-3}$ \\
 91.4 & 1.071$\times 10^{-3}$ & 4.498$\times 10^{-4}$ & 5.962$\times 10^{-5}$ &
 2.564$\times 10^{-3}$ & 1.463$\times 10^{-5}$ & 2.341$\times 10^{-3}$ \\
101   & 5.198$\times 10^{-4}$ & 2.183$\times 10^{-4}$ & 3.155$\times 10^{-5}$ &
1.357$\times 10^{-3}$ & 8.144$\times 10^{-6}$ & 1.303$\times 10^{-3}$ \\
111   & 2.655$\times 10^{-4}$ & 1.115$\times 10^{-4}$ & 1.752$\times 10^{-5}$ &
7.534$\times 10^{-4}$ & 4.749$\times 10^{-6}$ & 7.598$\times 10^{-4}$ \\
121   & 1.410$\times 10^{-4}$ & 5.922$\times 10^{-5}$ & 1.011$\times 10^{-5}$ &
4.347$\times 10^{-4}$ & 2.875$\times 10^{-6}$ & 4.600$\times 10^{-4}$ \\
131   & 7.736$\times 10^{-5}$ & 3.249$\times 10^{-5}$ & 6.025$\times 10^{-6}$ &
2.591$\times 10^{-4}$ & 1.795$\times 10^{-6}$ & 2.872$\times 10^{-4}$ \\
141   & 4.376$\times 10^{-5}$ & 1.838$\times 10^{-5}$ & 3.696$\times 10^{-6}$ &
1.589$\times 10^{-4}$ & 1.153$\times 10^{-6}$ & 1.845$\times 10^{-4}$ \\
151   & 2.539$\times 10^{-5}$ & 1.066$\times 10^{-5}$ & 2.323$\times 10^{-6}$ &
9.989$\times 10^{-5}$ & 7.589$\times 10^{-7}$ & 1.214$\times 10^{-4}$ \\
161   & 1.505$\times 10^{-5}$ & 6.321$\times 10^{-6}$ & 1.491$\times 10^{-6}$ &
6.411$\times 10^{-5}$ & 5.095$\times 10^{-7}$ & 8.152$\times 10^{-5}$ \\
171   & 9.078$\times 10^{-6}$ & 3.813$\times 10^{-6}$ & 9.743$\times 10^{-7}$ &
4.189$\times 10^{-5}$ & 3.482$\times 10^{-7}$ & 5.571$\times 10^{-5}$ \\
181   & 5.571$\times 10^{-6}$ & 2.340$\times 10^{-6}$ & 6.471$\times 10^{-7}$ &
2.783$\times 10^{-5}$ & 2.419$\times 10^{-7}$ & 3.870$\times 10^{-5}$ \\
191   & 3.478$\times 10^{-6}$ & 1.461$\times 10^{-6}$ & 4.369$\times 10^{-7}$ &
1.879$\times 10^{-5}$ & 1.707$\times 10^{-7}$ & 2.731$\times 10^{-5}$ \\
201   & 2.200$\times 10^{-6}$ & 9.240$\times 10^{-7}$ & 2.988$\times 10^{-7}$ &
1.285$\times 10^{-5}$ & 1.220$\times 10^{-7}$ & 1.952$\times 10^{-5}$ \\
\hline
\end{tabular}
\caption[]{The differential cross sections, in units of nb/GeV, 
without shadowing and rates in the interval
$|y_1| \leq 1$ in a one month, $10^6$ s, run
for resolved $\gamma+$jet photoproduction in peripheral $AA$ collisions.}
\label{comresrates}
\end{center}
\end{table}

\clearpage

\clearpage

\clearpage

\clearpage

\clearpage

\clearpage
\clearpage
\clearpage

\clearpage

\clearpage
\clearpage
\clearpage
\clearpage

\clearpage

\clearpage
\clearpage

\clearpage

\clearpage

\end{document}